# Hyperphosphorylation-Induced Phase Transition in Vesicle Delivery Dynamics of Motor Proteins in Neuronal Cells

Eunsang Lee[a, 1], Donghee Kim[b,c,d, 1], Yo Han Song[a, 1], Kyujin Shin[e], Sanggeun Song[b,c,d], Minho Lee[b,c,d], Yeongchang Goh[a], Mi Hee Lim[f], Ji-Hyun Kim[b], Jaeyoung Sung[b,c,d,*] and Kang Taek Lee[a,*]

**Affiliations:**

[a]Department of Chemistry, School of Physics and Chemistry, Gwangju Institute of Science and Technology (GIST), Gwangju 61005, Republic of Korea

[b]Creative Research Initiative Center for Chemical Dynamics in Living Cells, Chung-Ang University, Seoul 06974, Republic of Korea

[c]Department of Chemistry, Chung-Ang University, Seoul 06974, Republic of Korea

[d]National Institute of Innovative Functional Imaging, Chung-Ang University, Seoul 06974, Republic of Korea

[e]Materials Research & Engineering Center (MREC), R&D Division, Hyundai Motor Company, Uiwang 16082, Republic of Korea

[f]Department of Chemistry, Korea Advanced Institute of Science and Technology (KAIST), Daejeon 34141, Republic of Korea

* Jaeyoung Sung and Kang Taek Lee

**Email:** jaeyoung@cau.ac.kr and ktlee@gist.ac.kr

**Author Contributions:** [1] E.L., D.K., and Y.H.S. contributed equally to this work.

**Competing Interest Statement:** The authors declare no competing financial interest.

**Subject Areas:** Biological Physics, Physical Chemistry

**Abstract**

Synaptic vesicle transport by motor proteins along microtubules is a crucial active process underlying neuronal communication. It is known that microtubules are destabilized by tau-hyperphosphorylation, which causes tau proteins to detach from microtubules and form neurofibril tangles. However, how tau-phosphorylation affects transport dynamics of motor proteins on the microtubule remains unknown. Here, we discover that long-distance unidirectional motion of vesicle-motor protein multiplexes (VMPMs) in living cells is suppressed under tau-hyperphosphorylation, with the consequent loss of fast vesicle-transport along the microtubule. The VMPMs in hyperphosphorylated cells exhibit seemingly bidirectional random motion, with dynamic properties far different from VMPM motion in normal cells. We establish a parsimonious physicochemical model of VMPM's active motion that provides a unified, quantitative explanation and predictions for our experimental results. Our analysis reveals that, under hyperphosphorylation conditions, motor-protein-multiplexes have both static and dynamic motility fluctuations. The loss of the fast vesicle-transport along the microtubule can be a mechanism of neurodegenerative disorders associated with tau-hyperphosphorylation.

## I. INTRODUCTION

Neuronal communication is achieved by active transport of synaptic vesicles by kinesins and dyneins [1-7]. These molecular motors catalyze ATP-hydrolysis and convert the resulting chemical energy to mechanical energy required for their motion along the microtubule, a eukaryotic cytoskeleton formed by the polymerization of tubulin proteins. In the microtubule, tubulins are cross-linked by tau proteins to yield a stable helical structure, which produces the polarity of the microtubule [8-10]. This polarity allows kinesins and dyneins to move in the opposite direction from each other along the microtubule. Thus, a stable helical structure of the microtubule is essential for the microtubule-based directional motion of these motor proteins, which underlies many important biological processes, including the self-assembly and positioning of the mitotic spindle in cell division and the transport of various subcellular cargos ranging from vesicles to mitochondria [11-14].

Vesicles carried by these motor proteins exhibit unique transport dynamics far different from the dynamics of simple active matter motion or any kind of passive thermal motion. In living cells, multiple kinesins and dyneins are bound to a single vesicle and transport it along the microtubule [15,16]. In normal neuronal cells, this vesicle-motor protein multiplex (VMPM) exhibits multimode transport dynamics [17]. When the force on the vesicle exerted by kinesins in the anterograde direction (the direction towards the cell membrane) is in a delicate balance with the force exerted by dyneins in the retrograde direction (the direction towards the nucleus), the VMPM displays seemingly random bidirectional motion, as modeled by a tug-of-war [18-22]. The VMPM also shows unidirectional motion when either kinesin or dynein motors exert a far greater force on the vesicle [23,24], although the unidirectional motion of the VMPM is less frequent than the bidirectional motion [17]. At short times, fast liberation of the vesicle tethered to the motor protein multiplex (MPM) is dominant compared to the MPM motion along the microtubule. The resulting motion of the VMPM exhibits short-time sub-diffusion, transitioning to intermediate super-diffusion, and then ultimate diffusive motion [17,25]. This multimode transport dynamics of the VMPM in living cells was quantitatively explained by considering the reversible switching between unidirectional and bidirectional modes of the MPM as well as the liberational motion of the vesicle [17,26].

In cells under hyperphosphorylation conditions, the motion of VMPMs along the microtubule is expected to differ from their motion in normal cells. Hyperphosphorylated tau has been identified in many neurodegenerative diseases, including Alzheimer's disease, Down syndrome, and CBD [27]. In particular, the concentration of hyperphosphorylated tau proteins in the brains of Alzheimer's disease patients is found to be 3 to 4 times higher than that in normal adult brains [28,29]. Tau is a microtubule associated protein that plays a central role in nerve cells by maintaining cytoskeleton stability, regulating microtubule dynamics, promoting axon growth, and regulating axon migration [30,31]. Abnormally hyperphosphorylated tau proteins aggregate to form insoluble filaments (PHFs) and higher-order structures called intracellular neurofibrillary tangles (NTFs) [32-35]. NFTs are a primary marker of neurodegenerative disorders, such as Alzheimer's and Parkinson's; however, it is a controversial issue whether NFTs are their primary cause [36,37]. When hyperphosphorylated tau loses its affinity for microtubules and aggregates with tau oligomers in the cytoplasm, its ability to promote microtubule assembly or bind to microtubules is lost, resulting in the collapse of microtubules [32,38-43].

Despite extensive studies on the hyperphosphorylation of tau and the NFT formation and its physiological consequences [32,44], we do not yet know how tau-phosphorylation and ensuing tau-detachment from the microtubule affect the vesicle delivery dynamics of the VMPM along the microtubule. Tau in the microtubule can inhibit kinesin motion. A recent *in vitro* study showed that the speed of kinesin motion along the microtubule decreases with the concentration of tau in the microtubule [3,45-50]. This does not necessarily mean, however, that the VMPMs move faster upon hyperphosphorylation-induced tau detachment from the microtubule. So far, there is no quantitative understanding about the effects of tau-hyperphosphorylation on the VMPM motion along the microtubule.

To shed light on this issue, we monitored and analyzed individual trajectories of VMPMs along tau-deficient microtubules in living cells under hyperphosphorylation conditions. We treated human neuroblastoma cells with Forskolin, which results in a hyperphosphorylation-induced tau detachment from the microtubule [51]. Using upconverting nanoparticle (UCNP) probes free of photobleaching or photoblinking, we tracked nearly 450 individual trajectories of VMPMs in our cell system. Our investigation showed that, upon tau-hyperphosphorylation, unidirectional motion of the MPM is suppressed and the MPM effectively exhibits stochastic motion which appears Fickian-yet non-Gaussian diffusion along the microtubule in our experimental time resolution, 0.1 seconds. The observed transport dynamics of the VMPM is found to have qualitatively different stochastic properties from the previously reported thermal motion or active motion[52-54]. We propose a parsimonious physicochemical model of VMPM's active motion in hyperphosphorylated cells and perform a quantitative analysis of our experimentally monitored VMPM motion using the model. This simple model is found to provide a simultaneous explanation of the mean square displacement and the non-Gaussian parameter of the VMPM displacement distribution (VDD) and accurate predictions for the VDD in hyperphosphorylated cells at various time points. Our analysis shows the presence of both static and dynamic heterogeneity in motility of the MPMs in the forskolin-treated cells. The loss of fast synaptic vesicle transport along the microtubule upon tau-hyperphosphorylation can be a functional mechanism underlying diverse neurodegenerative disorders associated with tau-hyperphosphorylation [27].

## II. SYSTEM AND PROBE

Real-time trajectories of about 450 individual endosomal vesicles in a neuronal phenotype of human neuroblastoma SH-SY5Y cells under hyperphosphorylation conditions were collected (see Fig. 1 and the Supplemental Movie 1 for the real-time movie of a few UCNP containing vesicles moving along the axonal microtubule). To induce the hyperphosphorylation conditions, we treated our cells with 20uM of forskolin for 24 hours (see Appendix A). Hyperphosphorylation of tau protein in our cell system was confirmed by Western blotting [see Figs. 1(b) and 4]. As an imaging probe, we employed the upconverting nanoparticle (UCNP), free of photobleaching or photoblinking (see Appendix A 1 and Figs. 4 and 5), which enables accurate, long-time tracking of vesicle motion in living cells [55]. The conventional fluorescence probes have many important applications in the field of high-resolution microscopy and single molecule experiments [55-61]. However, for single vesicle tracking in living cells, our UCNP is advantageous over conventional fluorescence probes because of its outstanding photostability and low toxicity.

## III. EXPERIMENTAL RESULTS

### A. Primary direction of VMPM motion in hyperphosphorylated cells

The VMPM motion along the microtubule is far more pronounced than the motion in the microtubule-orthogonal direction in forskolin-treated cells as well as in normal cells [17] (see Figs. 7). This indicates that, even in our forskolin-treated cells, the microtubules are not broken apart and still play important roles in the microtubule-based transport of vesicles by motor proteins. We focused on VMPM motion in the microtubule direction in this work, unless stated otherwise.

### B. No preferred direction in VMPM motion along the microtubule

VMPM motion does not exhibit any bias between anterograde and retrograde directions along the microtubule. Individual VMPMs show stochastic motion biased in either the anterograde or retrograde direction; however, the mean displacement averaged over VMPMs is not biased to either of these two directions [see the inset shown in Fig. 1(d)]. This is also the case for normal cells [17]. Likewise, VMPM motion in the microtubule-orthogonal direction does not show any bias between upward and downward directions [see the inset presented in Fig. 1(e)].

### C. Unique stochastic properties of VMPM motion in hyperphosphorylated cells

The mean squared displacement (MSD) of VMPMs exhibits transient super-diffusive dynamics in normal cells, resulting from unidirectional MPM motion. In the forskolin-treated cells, however, the MSD does not exhibit such transient super-diffusive dynamics [Fig. 2(a)]. This observation indicates that, upon tau-hyperphosphorylation, unidirectional motion of MPM is suppressed.

Although active motion of the VMPMs appears to be bidirectional random motion in forskolin-treated cells, in stark contrast to active matter usually exhibiting unidirectional motion, their motion here has stochastic properties far different from thermal motion of molecules in complex fluids [62] or the Fickian yet non-Gaussian diffusion of colloidal beads along lipid tubes [63]. One of the most prominent features is that the long-time saturation of the non-Gaussian parameter (NGP) of the vesicle displacement to a constant value at our experimental time scale [Fig. 2(b)]. This is in contrast with the NGP time-profiles of molecules and colloids undergoing thermal motion in complex fluidic systems, which asymptotically decreases with time $t$ following $t^{-1}$ at long times [62]. The long-time saturation of the NGP value signifies the presence of static heterogeneity in VMPM motility or dynamic heterogeneity of VMPM motility whose relaxation time scale is longer than our observation time scale. The experimentally measured time-profiles of the MSD and NGP of VMPM motion in forskolin-treated cells cannot be explained by previously reported models.

## IV. THEORETICAL MODEL AND ANALYSIS

### A. Model and Theory of VMPM motion

To explain the MSD and NGP time-profiles of the observed VMPM motion, we should account for both VMPM motion along the microtubule and the liberational motion of a vesicle bound to the MPM. For this model, the displacement, $d_x$, of a vesicle can be represented by $d_x = R_x + x'$, where $R_x$ and $x'$ denote, respectively, the displacement of motor protein multiplex (MPM) and the change in the relative position of the vesicle with respect to the MPM. We assume that vesicle's librational motion around the MPM is so fast that $x'$ assumes a stationary distribution in our experimental time scale. Our experimental results shown in Fig. 2 tell us that the MPM motion in our forskolin-treated cells is non-Gaussian diffusion, which emerges when the motility of the MPM is distributed [62-71].

To explain our experiment, we construct a physicochemical model of the MPM motion in our system, in which MPMs alternatingly undergo unidirectional active motion and random changes in the direction of active motion. In this model, we assume that the MPM's motility is dependent on the state, $\Gamma$, of the MPM and the microtubule [62]. Using this model, we obtain the following analytic expressions of the MSD and NGP, $\langle d_x^2(t)\rangle$ and $\alpha_{2,x}(t)\left[\equiv \langle d_x^4\rangle \big/ \left(3\langle d_x^2\rangle^2\right) - 1\right]$, of the VMPM displacement at time scales longer than the time scale of random changes in the direction of MPM's active motion (see Appendix B 1)

$$\langle d_x^2(t)\rangle = 2\langle D_\Gamma\rangle t + \langle x'^2\rangle \tag{1a}$$

$$\alpha_{2,x}(t) = \frac{8}{\langle d_x^2(t)\rangle^2}\int_0^t dt'(t-t')\langle \delta D_\Gamma(t')\delta D_\Gamma(0)\rangle + \frac{\langle x'^2\rangle^2}{\langle d_x^2(t)\rangle^2}\alpha_2\left(x'\right), \tag{1b}$$

Where $\langle D_\Gamma\rangle$ and $\langle \delta D_\Gamma(t)\delta D_\Gamma(0)\rangle$ denote, respectively, the mean and the time-correlation-function (TCF) of the effective diffusion coefficient fluctuation of the MPM. In Eq. (1), $\langle x'^2\rangle$ and $\alpha_2\left(x'\right)$ designate the variance and NGP of the stationary distribution of vesicle's relative position with respect to the MPM (see Appendix B 8 and Fig. 8).

**B. Quantitative Analyses of MSD and NGP time-profiles**

The MSD time-profile of the VMPM in forskolin-treated cells is quantitatively explained by Eq. (1a) [Fig. 2(a)]. From this analysis, we extracted the average value of the effective diffusion coefficient of the MPM motion along the microtubule and the variance of the vesicle's relative position with respect to the MPM as $\langle D_\Gamma \rangle \cong 1.77\times10^{-2}\,\mu\text{m}^2/\text{s}$ and $\langle x'^2 \rangle = 0.0083\,\mu\text{m}^2$.

An accurate quantitative analysis of the NGP time-profile of VMPM motion is not an easy task. To analyze the NGP time-profile using Eq. (1b), we need a functional form of $\langle \delta D_\Gamma(t)\delta D_\Gamma(0)\rangle$, which depends on stochastic properties of the MPM motility fluctuation. It is often assumed that a TCF is a simple exponential function or a linear combination of a few exponential functions; however, this assumption is inconsistent with our experimental data, as discussed later. It is difficult to construct an accurate model at this stage due to the lack of information.

Instead of assuming a particular model of the MPM motility fluctuation from the beginning, we first extract the accurate time-profile of $\langle \delta D_\Gamma(t)\delta D_\Gamma(0)\rangle$ from our experimental data using

$$\left\langle \delta D(t)\delta D(0)\right\rangle = 8^{-1}\frac{d^2}{dt^2}\left[\left\langle d_x^2(t)\right\rangle^2 \alpha_{2,x}(t)\right], \tag{2}$$

which can be easily obtained from Eqs. (1a) and (1b). We emphasize that Eq. (2) enables us to extract the TCF of the MPM diffusion coefficient fluctuation directly from our experimental results for the MSD and NGP [see inset of Fig. 2(b)], without assuming a specific model of the MPM motility fluctuation. As shown in the inset of Fig. 2(b), $\langle \delta D_\Gamma(t)\delta D_\Gamma(0)\rangle$ extracted from our experimental results using Eq. (2) has a non-monotonic time-dependence and a finite long-time saturation value.

### C. Partially dynamic MPM motility fluctuation

The extracted time-profile of $\langle \delta D_\Gamma(t)\delta D_\Gamma(0)\rangle$ provides information useful for constructing a more explicit model of the MPM motility fluctuation. The long-time saturation of $\langle \delta D_\Gamma(t) \delta D_\Gamma(0) \rangle$ to a plateau value signifies that a group of MPMs has a static heterogeneity in the diffusion coefficient or dynamic heterogeneity whose relaxation occurs at times longer than our observation time [62]. The simple one-state model or the two-state dynamic model, $\Gamma_0 \rightleftarrows \Gamma_1$, of the MPM with a state-dependent motility cannot explain the long-time saturation of $\langle \delta D_\Gamma(t) \delta D_\Gamma(0) \rangle$ (Fig. 3). In contrast, any model assuming an entirely static distribution of the MPM motility yields $\langle \delta D_\Gamma(t) \delta D_\Gamma(0) \rangle$ constant in time, which does not exhibit a time-dependent relaxation.

We could quantitatively explain the time-profile of $\langle \delta D_\Gamma(t) \delta D_\Gamma(0) \rangle$ using a model that accounts for both static heterogeneity and dynamic heterogeneity in the MPM motility. A minimalistic, quantitative model is as follows. The MPM-microtubule system comprises two different groups. In one group, MPMs have a dynamically fluctuating motility, whose value changes over time depending on the MPM-microtubule state. In the other group, MPMs have a motility with negligible temporal fluctuations. A schematic representation of this model is given by

$$\Gamma_0 \underset{\psi_1(t)}{\overset{\psi_0(t)}{\rightleftarrows}} \Gamma_1 \qquad \Gamma_2, \tag{3}$$

where $\Gamma_0$, $\Gamma_1$, and $\Gamma_2$ represent the MPM-microtubule states, which will be simply designated by the MPM states from now on. Throughout, $D_j$ denotes the diffusion coefficient of the MPM at state $\Gamma_j$. Reversible transitions occur between states $\Gamma_0$ and $\Gamma_1$, but there is no transition to or from state $\Gamma_2$, at least in our experimental time scale. $\psi_0(t)$ and $\psi_1(t)$ denote the lifetime distribution of state $\Gamma_0$ and state $\Gamma_1$, respectively.

For the model in Eq. (3), the analytic expression of $\langle \delta D_\Gamma(t)\delta D_\Gamma(0)\rangle$ is given by

$$\langle \delta D_\Gamma(t)\delta D_\Gamma(0)\rangle = \sum_{i=0}^{1}\sum_{j=0}^{1} D_i G(\Gamma_i, t \mid \Gamma_j) D_j p_j + D_2^2 p_2 - \langle D_\Gamma\rangle^2 , \qquad (4)$$

where $p_j$ denotes the equilibrium probability of state $\Gamma_j$, satisfying the normalization condition, $\sum_{j=0}^{2} p_j = 1$ (see Appendix B 5). In Eq. (4), $G(\Gamma_i, t \mid \Gamma_j)$ denotes the probability that the MPM is at state $\Gamma_i$ at time *t*, given that the MPM is initially at state $\Gamma_j$. An explicit analytic expression of $G(\Gamma_i, t \mid \Gamma_j)$ is available as a functional of $\psi_0(t)$ and $\psi_1(t)$ [see Eq. (B11) in Appendix B 6] [72]. Eq. (1) with $\langle \delta D_\Gamma(t)\delta D_\Gamma(0)\rangle$ given by Eq. (4) provides a quantitative explanation of our experimental results for the time-profiles of the MSD and NGP of VMPM motion [Figs. 2(a) and 2(b)]. The optimized parameter values are given in Table 1 (see Appendix B 7).

**D. Prediction of the VMPM displacement distribution**

Using our optimized model of VMPM motion in the hyperphosphorylated cells, we predicted the spatial profile of the VMPM displacement distribution (VDD) (see Appendix B 8). The predictions of Eqs. (B15), (B16), and (B32) in Appendix B for our model are in quantitative agreement with the experimentally measured VDD at various times [Fig. 2(d)]. This remarkable agreement between theoretical prediction and experimental result for the VDD in our cell system shows that our model, which accounts for both static and dynamic motility fluctuations of the MPMs as well as the librational motion of a vesicle around each MPM, captures the essential features of the vesicle motion along the microtubule in hyperphosphorylated cells.

**V. DISCUSSION**

The hyperphosphorylation of tau suppresses the fast, long-distance vesicle transport of the motor proteins along the microtubule. This is because unidirectional motion of the MPM, which delivers vesicles faster over a long distance than bidirectional random motion, is negligible on the microtubule under tau-hyperphosphorylation. The suppression of unidirectional motion of the MPM is also manifested in vesicle trajectories in forskolin-treated cells (Fig. 7) and the spatial profile of VDD in forskolin-treated cells. Because the unidirectional motion causes the VDD to deviate from Gaussian, the VDD in normal cells has a strongly non-Gaussian heavy tail, which is absent in the VDD in forskolin-treated cells (Fig. 9) [17]. Together, these results clearly show that the fast, long-distance vesicle transport via unidirectional MPM motion is negligible in forskolin-treated cells.

A plausible mechanism of the hyperphosphorylation-induced change in MPM motion along the microtubule is a destabilization or structural change of the microtubule due to detachment of tau proteins from the microtubule upon hyperphosphorylation. Recent single molecule experiments clearly showed that tau proteins, which stabilize the helical structure of the microtubule, detach from the microtubule upon hyperphosphorylation [73,74], which is expected to deteriorate the structural stability of the microtubule. Other possible mechanisms of the change in the MPM motion include a structural change in the vesicle-MPM complex and a change in the dynamics of the ATP-hydrolysis catalyzed by the motor proteins. However, according to literature in this field, the hyperphosphorylation induced destabilization of the microtubule seems the primary cause of the loss of the change in the MPM motion on the microtubule upon hyperphosphorylation.

The loss of fast synaptic vesicle delivery by unidirectional MPM motion may be a mechanism of neurodegenerative disorders. The tau-hyperphosphorylation not only deteriorates the structural stability of neuronal cells [73,74], but also suppresses fast, long-distance delivery of synaptic vesicles by motor proteins along the microtubule, which would disrupt neuronal communications and brain functions. This may be a mechanism of neurodegenerative diseases associated with tau-hyperphosphorylation [75].

In our forskolin-treated cells, the mean diffusion coefficient of MPMs moving along the microtubule is estimated to be $\langle D_\Gamma \rangle \cong 1.77\times 10^{-2} \mu\text{m}^2/\text{s}$, far smaller than the diffusion coefficient value, $1.27\times 10^{-1} \mu\text{m}^2/\text{s}$, of dilute vesicles in cytoplasm [76]. On the other hand, the standard deviation of vesicle's relative position with respect to the MPM is $\sqrt{\langle x'^2 \rangle} \cong 0.091$ μm in forskolin-treated cells, which is about three-times greater than that in normal cells (Fig. 8). For this reason, VMPMs in forskolin-treated cells have greater MSD values at short times than VMPMs in normal cells [Fig. 2(a)].

We emphasize that Eq. (2) is applicable to various probe-attached systems exhibiting non-Gaussian diffusion, which enables one to extract the TCF of the system's motility fluctuation from the MSD and NGP of probe particle, without any prior knowledge or assumptions. By using this method, we estimate the normalized standard deviation, $\sigma_D / \langle D_\Gamma \rangle$, of the MPM diffusion coefficient to be about 2.1 and extract the non-monotonic time profile of the TCF, $\phi_D(t) \left[ \equiv \langle \delta D_\Gamma(t) \delta D_\Gamma(0) \rangle / \sigma_D^2 \right]$, of the diffusion coefficient fluctuation (inset of Fig. 2(b)). The time-profile of $\phi_D(t)$, or the NGP time profile, carries a lot more information than the MSD time profile. As shown in Fig. 3, the simple diffusion model of MPM motion, the dynamic two-state model, $\Gamma_0 \rightleftarrows \Gamma_1$, of the MPM with a state-dependent diffusion coefficient, and the static, three-state model of the MPM can all quantitatively explain the MSD time-profile of the VMPM, which obeys Eq. (1a). These models, however, cannot explain the time-profile of $\phi_D(t)$ or the NGP time-profile. In contrast, our MPM model with the partially dynamic MPM motility fluctuation, represented by Eq. (3), provides a simultaneous, quantitative explanation of the time profiles of $\phi_D(t)$, MSD, and NGP. According to our analysis, about 37% of MPMs exhibit temporal motility fluctuation while the remaining 63% of MPMs show negligible motility fluctuation (Table 1). We can think of more elaborated models of the MPM motion in hyperphosphorylated cells than ours. However, in this work, we choose arguably the simplest that quantitatively explains our experimental results. If we further simplify our model, we cannot explain the time-profiles of $\langle \delta D_\Gamma(t) \delta D_\Gamma(0) \rangle$ and the NGP (Fig. 3 and 10).

This work demonstrates an effective approach to quantitative investigation into dynamical processes occurring in complex systems. For a complex system, an experimental observable is often coupled with hidden dynamical variables about which we do not have much information. In this case, it is difficult to construct an explicit and accurate model from the beginning. However, if robust information about hidden variables coupled to our observable can be extracted by analyzing experimental data using a general theory, this information is useful for constructing a quantitative model of the complex system [77].

## VI. CONCLUSION

We shed lights on how tau-hyperphosphorylation affects the vesicle-motor-protein-multiplex motion along the microtubule in living cells. Upon tau-hyperphosphorylation in neuronal cells, the fast, long-distance cargo delivery via unidirectional MPM motion is suppressed. In our experimental time scale, MPMs exhibit a non-Gaussian diffusion due to partially dynamic motility fluctuations; about 37% of MPMs exhibit dynamic motility fluctuation and the remaining 63% static motility fluctuation in our experimental time window. The TCF of the diffusion coefficient fluctuation can be extracted from the MSD and NGP time profiles using Eq. (2); the TCF of the MPM diffusion coefficient fluctuation exhibits a unique, non-monotonic time-dependence. Using this information, we construct a minimalistic model that provides a simultaneous, quantitative explanation of the TCF of the MPM diffusion coefficient fluctuation, the MSD and NGP time-profiles of VMPM and even accurate predictions for the VMPM displacement distribution at various times for the VMPM moving along the microtubule in living cells under hyperphosphorylation conditions. The mean diffusion coefficient of the vesicle carried by MPMs on unstable microtubules is estimated to be $\langle D_\Gamma \rangle \cong 1.77\times10^{-2}\,\mu\mathrm{m}^2/\mathrm{s}$, only 14% of the diffusion coefficient of dilute vesicles undergoing thermal motion in cytoplasm. The normalized standard deviation, $\sigma_D/\langle D_\Gamma \rangle$, of the MPM diffusion coefficient is about 2.1. The standard deviation of the vesicle position with respect to the MPM is approximately 0.091 μm in forskolin-treated cells, about three-times greater than that in normal cells. The loss of fast cargo delivery dynamics may be a functional mechanism of various neurodegenerative diseases associated with tau-hyperphosphorylation.

**Acknowledgments**

K.T.L. was supported by the grant awarded from GIST in 2020 through the Research Institute (GRI) program and by the National Research Foundation (NRF) of S. Korea (Grant no. 2020R1F1A1073442, 2021R1A2C2010557). J.S. was supported by the Creative Research Initiative Project Program (NRF-2015R1A3A2066497) and the Engineering Research Center

Program funded by the Korea government (MSIT) (NRF-2020R1A5A1018052). Y.H.S., J.-H.K., and M.H.L. were supported by the NRF (NRF-2019R1I1A1A01056975 (Y.H.S.); NRF-2020R1A2C1102788 (J.-H.K.); NRF-2022R1A3B1077319 (M.H.L.)).

## APPENDIX A: EXPERIMENTAL MATERIALS AND METHODS

### 1. Cell system and imaging probe

SH-SY5Y cells were differentiated to neuronal phenotype by retinoic acid treatment (see Appendix A 2). Then, the differentiated cells were treated by Forskolin to induce hyperphosphorylation of Tau. We confirm hyperphosphorylation of tau protein in our cells by Western blotting to estimate pTau-S214 and total tau levels (see Appendix A 3). As the imaging probe, we adopted $Yb^{3+}$, $Er^{3+}$ doped core/shell UCNPs ($NaY_{0.78}F_4$:$Yb^{3+}_{0.2}$, $Er^{3+}_{0.02}$@$NaYF_4$@PAA). These probes have hexagonal structure and a 24.5 (±1.0) nm size (Fig. 5) and have green (530 nm and 550 nm) and red emissions (650 nm) under 980 nm excitation [Fig. 6(a)]. For simultaneous, real-time tracking of multiple UCNP probes, we use the wide-field based epi-fluorescence microscope developed in our previous work [Fig. 6(b)] [17].

### 2. Cell culture

SH-SY5Y neuroblastoma cells (Korea Cell Line Bank) were cultured in Dulbecco's Modified Eagle's Medium (DMEM, Hyclone) with 10% fetal bovine serum (FBS, Hyclone), 100 units/mL penicillin, and 100 g/mL streptomycin in a 10 mL culture dish and incubated at 37°C under humidified 5% $CO_2$. Once the cells reached 70-80% confluency, they were treated with trypsin-EDTA to obtain the cell suspension. 1 mL of this suspension ($1\times10^4$ cells/mL) was seeded on a poly-D-Lysine (Sigma-Aldrich)-treated cover glass bottomed dish (SPL) and incubated for 24 h at 37°C in an incubator humidified with 5% $CO_2$. For cell differentiation, 10 mM retinoic acid (Sigma-Aldrich) was added to the cultured cells. Differentiation was allowed to proceed for more than 10 days and the media was exchanged regularly (every two days) with freshly prepared media containing 10 mM retinoic acid. After that, 20 μM of forskolin was treated to the differentiated cells for 24 hours for inducing hyper-phosphorylation of tau proteins in the cell.

Prior to imaging, 2 mg/mL of upconverting nanoparticles (UCNPs) (in PBS) was treated to differentiated SH-SY5Y cells to a final concentration of 20 mg/mL, and incubated (37°C, 5% $CO_2$) for 30 minutes for intracellular uptake. After incubation, cells were washed 3 times with PBS, filled with imaging medium, mounted in a live cell chamber (TC-L-10, Live Cell Instrument) maintained at 37°C 5% $CO_2$.

Then, UCNPs in these differentiated cells were imaged using a wide-field based inverted microscope (IX73, Olympus) under 980 nm diode laser (P151-500-980A, EM4 Inc.) irradiation with a 200 mW/cm$^2$ pump power for 30 min [Fig. 6(b)] and images of the endosomal vesicles containing UCNPs were obtained by EMCCD camera (iXON3, Andor Technology). The images of cellular shapes were acquired under bright-field lamp irradiation. For acquisition of cell and UCNPs images, Solis software (Andor Technology) was used (The exposure time for the camera was set typically at 50 ms). After that, 450 trajectories, at least 20 s long, were obtained from the series of such-set images by using particle tracking software (DiaTrack).

**3. Western blotting**

To carry out the Western blot, SH-SY5Y cells were differentiated in a 10 mL culture dish and treated with 20 μM of Forskolin prepared in DMSO for 24 hours. Simultaneously, a blank experiment was run where only DMSO was used. Cell lysate was obtained by treatment with lysis buffer containing the inhibitors phosphatase and protease. SDS-PAGE was performed using NovexTM Value 4-20% Tris-Glycine Gel (Invitrogen), and the protein bands were transferred to Immobilon-P membrane (Milipore). Thereafter, the primary antibody (Ser-214, total Tau, β-actin; Abcam) reaction was carried out followed by the HRP-labeled secondary antibody reaction. The color image was visualized and quantified with a phosphorimager (Chemidoc, Biorad) and software.

In Ref. [51], it was shown that the forskolin-induced hyperphosphorylation of tau protein leads to the destabilization of the microtubule. Forskolin activates protein kinase A by stimulating adenylyl cyclase that increases intracellular levels of cAMP [78,79]. With our western blot experiment, we confirm that phosphorylation of Ser-214 site by the forskolin treatment. Ser-214 is one of the phosphorylation sites of tau protein found in AD patients [80-89].

## 4. Chemical

For the synthesis of $\beta$–$NaY_{0.78}F_4$: $Yb^{3+}_{0.2}$, $Er^{3+}_{0.02}$ @$NaYF_4$ core/shell nanoparticles, yttrium (III) acetate hydrate (99.9%), ytterbium (III) acetate tetrahydrate (99.9%), erbium (III) acetate hydrate (99.9%), 1-octadecene (technical grade, 90%), oleic acid (technical grade, 90%), ammonium fluoride (≥98%), and sodium hydroxide (≥98%), poly(acrylic acid) (PAA) were sourced from Sigma-Aldrich. For cell culture and differentiation, retinoic acid purchased from Sigma-Aldrich and Dulbecco's Modified Eagle's Medium (DMEM) and 10 % fetal bovine serum (FBS) were sourced from Hyclone. All chemical reagents were used without further purification.

**5. Synthesis of $\beta$–$NaY_{0.78}F_4$: $Yb^{3+}_{0.2}$, $Er^{3+}_{0.02}$ core nanoparticles.**

$Y(CH_3COO)_3{\cdot}xH_2O$ (0.78 mmol), $Yb(CH_3COO)_3{\cdot}xH_2O$ (0.2 mmol), and $Er(CH_3COO)_3{\cdot}xH_2O$ (0.02 mmol) were added to a 100 mL three-neck round-bottom flask containing oleic acid (6 mL) and 1-octadecene (15 mL). Subsequently, the solution was heated to 150 ℃ with stirring for 1 hour and then cooled to room temperature. A mixture of ammonium fluoride (4.0 mmol) and sodium hydroxide (2.5 mmol) dissolved in methanol (10 mL) was added to the reaction flask and stirred for 1 hour at 50 ℃. After that, the reaction mixture was heated to 100 ℃ under vacuum with stirring for 15 min to remove remaining methanol. Then, the reaction mixture was heated to 300 ℃ and kept at this temperature for 1 h under Ar gas. After cooling down to room temperature, UCNPs were purified by repetitive centrifugation and washing.

**6. Synthesis of $\beta$–$NaY_{0.78}F_4$: $Yb^{3+}_{0.2}$, $Er^{3+}_{0.02}$ @$NaYF_4$ core/shell nanoparticles.**

The synthesis method of core/shell nanoparticles is almost the same as that of core nanoparticles. $Y(CH_3COO)_3{\cdot}xH_2O$ (0.5 mmol) was added to a 100 mL three-neck round-bottom flask containing oleic acid (6 mL) and 1-octadecene (15 mL). After heating to 150 ℃ with stirring for 1 hour and cooling to room temperature, a solution of hexagonal phase core UCNPs in hexane (1 mmol) synthesized in the previous process was added to the reaction flask. Then, the mixture was heated to 80 ℃ with stirring for 1 hour to remove hexane, and then cooled to room temperature. A mixture of ammonium fluoride (2.0 mmol) and sodium hydroxide (1.25 mmol) dissolved in methanol (5 mL) was added to the reaction flask and stirred for 30 min at 50 ℃. Subsequently, the reaction mixture was heated to 100 ℃ under vacuum with stirring for 15 min to remove methanol. After that, the reaction mixture was heated to 300 ℃ and kept at this temperature for 1 hour under Ar gas. After cooling down to room temperature, UCNPs were purified by repetitive centrifugation and washing. Their TEM image is shown in Fig. 5(a).

**7. Surface modification of $\beta$–$NaY_{0.78}F_4$: $Yb^{3+}_{0.2}$, $Er^{3+}_{0.02}$ @$NaYF_4$ core/shell nanoparticles.**

The surfaces of synthesized core/shell UCNPs were modified with $NOBF_4$ (nitrosyl tetrafluoroborate). 2 mL of DMF was added to 50 mg of core/shell UCNPs in hexane. And then, 50 mg of $NOBF_4$ was added into the DMF/UCNPs solution. After stirring for 15 min with 800 rpm, the solution was centrifuged at 4000 rpm for 5 min. After removing the supernatant, the nanoparticles were dispersed in 10 mL DMF.

The surfaces of core/shell UCNPs with $NOBF_4$ were modified with PAA. 0.6 mL of 2.0 M NaOH solution, 1.4 mL of 3-D water, and 100 mg of PAA were mixed, and the solution was stirred at 1200 rpm. The 5 mg of prepared core/shell UCNPs with $NOBF_4$ was added into the solution dropwise. After 30 min, the UCNPs-PAA were purified by repetitive centrifugation and washing three times with 3-D water. Their TEM image is shown in Fig. 5(b).

## APPENDIX B: THEORETICAL METHODS

### 1. MSD and NGP of the VMPM

The displacement of the MPM and the change in the relative motion of the vesicle with respect to the MPM are not strongly correlated. Therefore, the second and fourth moments of vesicle displacement are given by

$$\langle d_x^2(t)\rangle = \langle R_x^2(t)\rangle + \langle x'^2\rangle, \qquad \text{(B1a)}$$

$$\langle d_x^4(t)\rangle = \langle R_x^4(t)\rangle + \langle x'^4\rangle + 6\langle R_x^2(t)\rangle\langle x'^2\rangle, \qquad \text{(B1b)}$$

where $\langle R_x^n(t)\rangle$ and $\langle x'^n\rangle$ denote, respectively, the $n$-th moment of the MPM displacement distribution and the $n$-th moment of the distribution of vesicle's relative displacement with respect to the MPM. The MSD of the VMPM designates $\langle d_x^2(t)\rangle$ given in Eq. (B1a).

Substituting Eqs. (1a) and (1b) into the definition of the non-Gaussian parameter (NGP), $[\langle d_x^4(t)\rangle / 3\langle d_x^2(t)\rangle^2] - 1$, we obtain a general expression of the NGP of the VMPM displacement,

$$\alpha_{2,x}(t) = \frac{\langle R_x^2(t)\rangle^2}{\langle d_x^2(t)\rangle^2}\alpha_2\left(R_x, t\right) + \frac{\langle x'^2\rangle^2}{\langle d_x^2(t)\rangle^2}\alpha_2\left(x'\right), \qquad \text{(B2)}$$

where $\alpha_2(R_x, t)$ and $\alpha_2(x')$ represent, respectively, the NGP of the $R_x$ distribution and the NGP of the $x'$ distribution. In Eqs. (B1) and (B2), because $x'$ assumes a stationary distribution in our experimental time resolution, $\langle x'^2\rangle$ and $\alpha_2(x')$ are set to be constant in time. On the other hand, the displacement of the MPM motion along the microtubule has a non-stationary distribution, so the values of $\langle R_x^2(t)\rangle$ and $\alpha_2(R_x, t)$ change over time. We emphasize that Eqs. (B1) and (B2) are applicable not only to our VMPM system but also to any probe-attached system regardless of the system's transport dynamics. However, to analyze the NGP time profile using Eq. (B2), we need an explicit functional form of $\langle R_x^2(t)\rangle$, which is dependent on a model of MPM motion along the microtubule.

**2. Transport equation describing MPM motion along the microtubule in hyperphosphorylated cells**

In our model, the MPM alternatingly undergoes unidirectional active motion and a random change in the direction of the active motion where the speed of the unidirectional MPM motion is dependent on the MPM state, $\Gamma$, which includes the microtubule state. At time scales far longer than the time scale of the random change in the direction of the MPM motion, the probability distribution of the position and state of MPMs can be described by

$$\hat{\dot{p}}(x,\mathbf{\Gamma},s) = D_{\Gamma}\frac{\partial^2}{\partial x^2}\hat{p}(x,\mathbf{\Gamma},s) + L(\mathbf{\Gamma})\hat{p}(x,\mathbf{\Gamma},s)\,, \qquad \text{(B3)}$$

where $p(x,\mathbf{\Gamma},t)$ designates the joint probability density that an MPM is located at position $x$ and the MPM is at state $\mathbf{\Gamma}$ at time $t$, satisfying the normalization condition: $\int_{-\infty}^{\infty} dx \int d\mathbf{\Gamma}\, p(x,\mathbf{\Gamma},t) = 1$. In Eq. (B3), $\hat{p}(x,\mathbf{\Gamma},s)$ and $\hat{\dot{p}}(x,\mathbf{\Gamma},s)$ denote the Laplace transforms of $p(x,\mathbf{\Gamma},t)$ and $\partial_t p(x,\mathbf{\Gamma},t)$ with $s$ being the Laplace variable. $D_{\Gamma}$ and $L(\mathbf{\Gamma})$ in Eq. (B3) denote, respectively, the effective diffusion coefficient of the MPM at state $\mathbf{\Gamma}$ and a mathematical operator describing dynamics of MPM state $\mathbf{\Gamma}$. At this stage, we do not have information about the MPM-state dynamics required to construct an explicit mathematical form of operator $L(\mathbf{\Gamma})$. However, by analyzing our experimental results using a general solution of Eq. (B3), we obtain the analytic expressions for the MSD and NGP of the VMPM and extract information about the MPM state dynamics and the mathematical form of $L(\mathbf{\Gamma})$, as demonstrated in the main text. Equation (B3) can also be derived by considering the continuum limit of an unbiased random walk of the MPM, in which the speed of the unidirectional motion of the MPM to the nearest neighbor site on the microtubule is not a constant but a function of MPM state $\mathbf{\Gamma}$ (see Appendix B 11). We also note there that Eq. (B3) is a limiting form of the general transport equation describing non-Fickian and non-Gaussian motion in complex systems[62,90], which can be derived either by considering the environmental state dependent continuous time random walk model or by applying the projection operator technique to a general many-particle system that obeys Newton's classical dynamics [91,92].

### 3. Mean square displacement and the non-Gaussian parameter of MPM motion

From Eq. (B3), we obtain the general analytic expression for the second and fourth moments, $\langle R_x^2(t)\rangle (\equiv \langle [R_x(t)-R_x(0)]^2\rangle)$ and $\langle R_x^4(t)\rangle (\equiv \langle [R_x(t)-R_x(0)]^4\rangle)$, of the MPM displacement using a similar mathematical method as Ref. [62]. The analytic results are given by

$$\langle R_x^2(t)\rangle = 2\langle D_\Gamma\rangle t\,, \tag{B4a}$$

$$\langle R_x^4(t)\rangle = 3\langle R_x^2(t)\rangle^2 + 24\int_0^t d\tau (t-\tau)\langle \delta D_\Gamma(\tau)\delta D_\Gamma(0)\rangle\,, \tag{B4b}$$

where $\langle D_\Gamma\rangle$ and $\langle \delta D_\Gamma(\tau)\delta D_\Gamma(0)\rangle$ denotes, respectively, the mean and the time correlation function (TCF) of the diffusion coefficient fluctuation of the MPM moving along the microtubule in hyperphosphorylated cells.

Substituting Eqs. (B4a) and (B4b) into the definition of the NGP, $\alpha_2(R_x,t)[\equiv \langle R_x^4\rangle / (3\langle R_x^2(t)\rangle^2)-1]$, of the MPM displacement, we obtain

$$\alpha_2(R_x,t) = 2\eta_D^2 t^{-2}\int_0^t d\tau (t-\tau)\phi_D(\tau)\,, \tag{B5a}$$

where $\eta_D^2$ and $\phi_D(\tau)$ denotes the relative variance and the normalized TCF of the diffusion coefficient fluctuation, defined by $\eta_D^2 \equiv \langle \delta D_\Gamma^2\rangle / \langle D_\Gamma\rangle^2$ and $\phi_D(\tau) = \langle \delta D_\Gamma(\tau)\delta D_\Gamma(0)\rangle / \langle \delta D_\Gamma^2\rangle$. Substituting Eqs. (B4a) and (B5a) into the Eqs. (B1) and (B2), we obtain Eq. (1) in the main text. At short times where $\phi_D(\tau)\cong 1$, the NGP given in Eq. (B5a) becomes approximately the same as $\eta_D^2$, i.e., $\alpha_2(R_x,t)\cong \eta_D^2$. In contrast, at long times, the NGP has the following asymptotic

behavior:

$$\alpha_2(R_x,t) \cong \eta_D^2 \phi_D(\infty) + 2\eta_D^2 \, t_D/t \,, \tag{B5b}$$

where $t_D$ is the relaxation time of the diffusion coefficient fluctuation, defined by $t_D \equiv \int_0^\infty d\tau [\phi_D(\tau) - \phi_D(\infty)]$.

The long-time limit value of $\alpha_2(R_x,t)$ is finite only when $\phi_D(t)$ does not vanish at long times. From Eq. (B5b), one can see that $\lim_{t\to\infty} \alpha_2(R_x,t) = \phi_D(\infty)\eta_D^2$, likewise one can obtain asymptotic behavior of the non-Gaussian parameter of the vesicle-motor protein multiplex (see Appendix B 4). $\phi_D(\infty)$ has a finite value only when the MPM motility depends on the MPM state and there exists an MPM state or a group of MPM states from or to which no transition occurs.

**4. Asymptotic behaviors of the non-Gaussian parameter**

Substituting Eqs. (B4a) and (B5a) into the Eq. (B2), we obtain an analytic expression of the NGP of the VMPM displacement,

$$\alpha_{2,x}(t) = \frac{8\langle \delta D_\Gamma^2 \rangle}{\langle d_x^2(t) \rangle^2} \int_0^t dt'(t-t')\phi_D(t') + \frac{\langle x'^2 \rangle^2}{\langle d_x^2(t) \rangle^2} \alpha_2\left(x'\right) \tag{B6}$$

At short times, $\langle d_x^2(t) \rangle$ given in Eq. (B6) can be approximated by $\langle d_x^2(t) \rangle \cong \langle x'^2 \rangle$, and the value of $\phi_D(t) \left(= \langle \delta D_\Gamma(t) \delta D_\Gamma(0) \rangle / \langle \delta D_\Gamma^2 \rangle \right)$ is approximately equal to unity. With these approximations at hand, Eq. (B6) can be approximately written as $\alpha_{2,x}(t) \cong 4t^2 \langle \delta D_\Gamma^2 \rangle / \langle x'^2 \rangle^2 + \alpha_2(x')$. Therefore, in the short time limit, we have $\lim_{t\to 0} \alpha_{2,x}(t) = \alpha_2(x')$. That is to say, at short times, the NGP of the vesicle-MPM displacement is primarily determined by the NGP of the distribution of the relative position of the vesicle with respect to MPM. On the other hand, at long times, $\langle d_x^2(t)\rangle$ given in Eq. (1a) can be approximated by $\langle d_x^2(t)\rangle \cong 2\langle D_\Gamma\rangle t$. Given that the value of $\phi_D(t)$ approaches $\phi_D(\infty)$ at long times, Eq. (B6) becomes $\alpha_{2,x}(t) \cong \langle \delta D_\Gamma^2\rangle / \langle D_\Gamma\rangle^2 \phi_D(\infty) + t^{-2}\langle x'^2\rangle^2 \alpha_2(x') / (4\langle D_\Gamma\rangle^2)$. Therefore, in the long-time limit, we have $\lim_{t\to\infty} \alpha_{2,x}(t) = \phi_D(\infty)\eta_D^2$ with $\eta_D^2$ being $\langle \delta D_\Gamma^2\rangle / \langle D_\Gamma\rangle^2$.

**5. Analytic expression for the TCF of the diffusion coefficient fluctuation**

The time-correlation function of the diffusion coefficient is defined by

$$\langle D_\Gamma(t) D_\Gamma(0)\rangle = \int d\Gamma \int d\Gamma_0 D_\Gamma G(\Gamma, t \mid \Gamma_0) D_{\Gamma_0} p_I(\Gamma_0), \tag{B7}$$

where $G(\Gamma, t \mid \Gamma_0)$ denotes the probability that the MPM state is found at $\Gamma$ at time *t*, given that the MPM is initially at state $\Gamma_0$. $p_I(\Gamma_0)$ denotes the probability that the MPM is initially at state $\Gamma_0$. The TCF of the diffusion coefficient fluctuation, $\delta D_\Gamma(t) = D_\Gamma(t) - \langle D_\Gamma\rangle$, is given by

$$\langle \delta D_\Gamma(t) \delta D_\Gamma(0)\rangle = \langle D_\Gamma(t) D_\Gamma(0)\rangle - \langle D_\Gamma\rangle^2. \tag{B8}$$

For our discrete MPM state model given in Eq. (3), Eqs. (B7) and (B8) yield Eq. (4) in the main text, and the analytic expression of $G(\Gamma_i, t \mid \Gamma_j)$ in the Laplace domain is given in Appendix B 6, Eq. (B11)

**6. Analytic expression of $\phi_D(t)$**

Our model yields an analytic expression for the time correlation function of the diffusion coefficient, which is defined by

$$\langle D_{\Gamma}(t)D_{\Gamma}(0)\rangle = \sum_{i=0}^{2} D_i G(\Gamma_i, t \mid \Gamma_j) D_j p_j \,, \tag{B9}$$

where $p_j$ denotes the equilibrium probability of state $\Gamma_j$, satisfying the normalization condition, $\sum_{j=0}^{2} p_j = 1$. In Eq. (B9), $G(\Gamma_i, t \mid \Gamma_j)$ denotes the propagator or the conditional probability of finding the system at state $\Gamma_i$ at time *t*, given that our system is at state $\Gamma_j$ at time 0. Since there is no transition to or from state $\Gamma_2$ in our model, Eq. (B9) becomes

$$\langle D_{\Gamma}(t)D_{\Gamma}(0)\rangle = \sum_{i=0}^{1} D_i G(\Gamma_i, t \mid \Gamma_j) D_j p_j + D_2^2 p_2 . \tag{B10}$$

$G(\Gamma_i, t \mid \Gamma_j)$ in Eq. (B10) is simply related to the lifetime distributions, $\psi_0(t)$ and $\psi_1(t)$, of states $\Gamma_0$ and $\Gamma_1$ in the Laplace transform domain. Given that $\mathbf{G}(t)$ denotes the matrix whose element is given by $G(\Gamma_i, t \mid \Gamma_j)$, its Laplace transform is given by[72]

$$\hat{\mathbf{G}}(s) = \frac{1}{s}\begin{pmatrix} 1 - \dfrac{1}{\langle t_0\rangle s}\dfrac{\left(1-\hat{\psi}_0(s)\right)\left(1-\hat{\psi}_1(s)\right)}{1-\hat{\psi}_1(s)\hat{\psi}_0(s)} & \dfrac{1}{\langle t_1\rangle s}\dfrac{\left(1-\hat{\psi}_0(s)\right)\left(1-\hat{\psi}_1(s)\right)}{1-\hat{\psi}_1(s)\hat{\psi}_0(s)} \\ \dfrac{1}{\langle t_0\rangle s}\dfrac{\left(1-\hat{\psi}_0(s)\right)\left(1-\hat{\psi}_1(s)\right)}{1-\hat{\psi}_1(s)\hat{\psi}_0(s)} & 1 - \dfrac{1}{\langle t_1\rangle s}\dfrac{\left(1-\hat{\psi}_0(s)\right)\left(1-\hat{\psi}_1(s)\right)}{1-\hat{\psi}_1(s)\hat{\psi}_0(s)} \end{pmatrix}. \tag{B11}$$

$p_0$ and $p_1$ in Eq. (B10) are related to the mean lifetimes, $\langle t_0\rangle$ and $\langle t_1\rangle$, of state $\Gamma_0$ and $\Gamma_1$, satisfying the normalization condition, $p_0 + p_1 = 1 - p_2$, i.e.,

$$p_0 = \frac{\langle t_0\rangle}{\langle t_0\rangle + \langle t_1\rangle}\left(1 - p_2\right), \tag{B12a}$$

$$p_1 = \frac{\langle t_1 \rangle}{\langle t_0 \rangle + \langle t_1 \rangle}\left(1 - p_2\right), \qquad \text{(B12b)}$$

Using Eqs. (B10)-(B12), we calculate the Laplace transform, $\hat{\phi}_D(s)$, of the normalized time correlation function of the diffusion coefficient fluctuation, defined by

$$\phi_D(t) = \frac{\langle \delta D_\Gamma(t)\delta D_\Gamma(0)\rangle}{\langle \delta D_\Gamma^2(0)\rangle} = \frac{\langle D_\Gamma(t) D_\Gamma(0)\rangle - \langle D_\Gamma \rangle^2}{\langle D_\Gamma^2 \rangle - \langle D_\Gamma \rangle^2} \quad . \qquad \text{(B13)}$$

Where $\langle D_\Gamma^k \rangle = \sum_{i=0}^{2} D_i^k p_i$ . Using Eqs. (B6) and (B13), we have quantitatively analyzed the experimental results of the NGP time profile of the vesicle-MPM displacement. To analyze the time profiles of the MSD and NGP data using our results in the Laplace transform domain, we performed numerical inverse Laplace transform of our theoretical results using Stehfest algorithm [93]. The best-fitted result of our theory is compared with the experimental results for the NGP in Fig. 2(b). From the quantitative analysis, we extracted the optimized values of the adjustable parameters for the vesicle-MPM motion along the microtubule. These values are collected in Table 1.

**7. Quantitative model of MPM motion in hyperphosphorylated cells.**

The time-profile of $\langle \delta D_\Gamma(t)\delta D_\Gamma(0)\rangle$ extracted from our analysis of experimental data is quantitatively explained by Eq. (4), obtained for the model of MPM motion in Eq. (3), if the MPM state lifetime distributions, $\psi_0(t)$ and $\psi_1(t)$, are assumed to be gamma distributions (Fig. 11). The mean and relative variance of the MPM state lifetimes are extracted as $\langle t_0 \rangle \cong 2.85$s and $\langle \delta t_0^2 \rangle / \langle t_0 \rangle^2 \cong 0.14$ for state $\Gamma_0$ and $\langle t_1 \rangle \cong 1.34$s and $\langle \delta t_1^2 \rangle / \langle t_1 \rangle^2 \cong 0.74$ for state $\Gamma_1$. According to these results, the transitions between the MPM states are sub-Poisson processes for which the relative variance of the MPM state lifetime is less than unity. The sub-Poisson state transition dynamics emerges when the state transition is a multi-step process [94]. The non-monotonic time dependence of the normalized TCF of the diffusion coefficient fluctuation, $\phi_D(t)\left[\equiv\langle\delta D_\Gamma(t)\delta D_\Gamma(0)\rangle/\langle\delta D_\Gamma^2\rangle\right]$, shown in inset of Fig. 2(b), results from the sub-Poisson MPM state transition processes. $\langle\delta D_\Gamma(t)\delta D_\Gamma(0)\rangle$ would have monotonically decreased with time if the transitions between the MPM states were Poisson processes or first-order kinetic processes with constant rate coefficients (Fig. 10) or if they were super-Poisson processes or multi-channel processes.

**8. Distribution of a vesicle's position relative to the MPM carrying the vesicle**

The relative position of a vesicle with respect to the MPM assumes a stationary distribution in our experimental time resolution, 0.1 seconds. We obtain the stationary distribution of vesicle position with respect to the MPM from the VDD at short times at which MPM motion is negligible (Fig. 8). The experimentally observed stationary distribution of vesicle around the MPM can be explained by the following superposition of Gaussians with heterogeneous variance [17]:

$$f(x')=\int_{-\infty}^{\infty}dq\sqrt{\frac{\kappa_0}{2\pi(1+q^2)}}\exp\left(-\frac{\kappa_0 x'^2}{2(1+q^2)}\right)N(q\,|\,\sigma_q^2), \tag{B14}$$

where $\kappa_0$ and $\sigma_q^2$ are constant (Fig. 8). In Eq. (B14), $N(q\,|\,\sigma_q^2)$ denotes the normal distribution of $q$ with zero mean and variance $\sigma_q^2$. The optimized parameter values are given by $\kappa_0^{-1}=2.4\times10^{-3}\,\mu\mathrm{m}^2$ and $\sigma_q^2=2.46$. The variance, $\langle x'^2\rangle$, and NGP, $\alpha_2(x')$, are given by $\kappa_0^{-1}(1+\sigma_q^2)\left(\cong 0.0083\mu\mathrm{m}^2\right)$ and $2\sigma_q^4\big/\left(1+\sigma_q^2\right)^2\left(\cong 1.0096\right)$ respectively. The Fourier transform of Eq. (B14) is given by

$$\tilde{f}(k)=\exp\left(-k^2/2\kappa_0\right)\Big/\sqrt{\sigma_q^2\left(k^2/\kappa_0\right)+1}\,. \tag{B15}$$

with $k$ denoting the Fourier variable, i.e., $\tilde{f}(k)=\int_{-\infty}^{\infty}dx'e^{-ikx'}f(x')$.

## 7. Distribution of the MPM displacement

For our MPM model given in Eq. (3), we obtain the analytic expression for the MPM displace distribution (see Appendix B 8) in the Fourier-Laplace transform domain, which read as

$$\hat{\tilde{P}}(k,s)=-\left(1-p_2\right)\frac{\left[1-\hat{\psi}_0(s_0)\right]\left[1-\hat{\psi}_1(s_1)\right]}{\left(\langle t_0\rangle+\langle t_1\rangle\right)\left[1-\hat{\psi}_0(s_0)\hat{\psi}_1(s_1)\right]}\left(\frac{1}{s_0}-\frac{1}{s_1}\right)^2+\frac{p_0}{s_0}+\frac{p_1}{s_1}+\frac{p_2}{s_2}\,, \tag{B16}$$

where $\hat{\tilde{P}}(k,s)$ designates the Fourier-Laplace transform of the MPM displacement distribution, $P(R_x,t)$, i.e., $\hat{\tilde{P}}(k,s)=\int_{-\infty}^{\infty}dR_x\exp(ikR_x)\int_0^{\infty}ds\exp(-st)P(R_x,t)$ and $s_i$ denotes $s+D_ik^2$ with $s$ and $k$ denoting the Laplace variable and the Fourier variable, respectively In Eq. (B16), $p_i$ denotes the equilibrium probability of MPM state $\Gamma_i$, which satisfies $\sum_{i=0}^{2}p_i=1$, and $p_0$ and $p_1$ are dependent on the mean lifetimes of $\Gamma_0$ and $\Gamma_1$ and $p_2$, i.e., $p_0=(1-p_2)\langle t_0\rangle\big/\left(\langle t_0\rangle+\langle t_1\rangle\right)$ and $p_1=(1-p_2)\langle t_1\rangle\big/\left(\langle t_0\rangle+\langle t_1\rangle\right)$ with $\langle t_0\rangle$ and $\langle t_1\rangle$ being the mean lifetime of MPM state 0 and 1, that is, $\langle t_i\rangle=\int_0^{\infty}dt\;t\psi_i(t)$.

## 9. Analytic expression for $\hat{\tilde{P}}(k,s)$

In this section, we provide a derivation of Eq. (B16). First, we note that, under the stationary initial condition in which we start our observation of the system at a randomly chosen time between state transition events, the probability density, $\psi_j^{st}(t)$, of the time at which the first state transition occurs is given by[95]

$$\psi_j^{st}(t) = \frac{S_j(t)}{\langle t_j \rangle}, \tag{B17a}$$

where $S_j(t)$ denotes the survival probability, defined by $S_j(t) = \int_t^\infty \psi_j(\tau) d\tau$ with $\psi_j(t)$ being the lifetime distribution of our system at state $\Gamma_j$. The probability that the first state transition does not occur as of time *t* since our observation is given by

$$S_j^{st}(t) = \int_t^\infty \psi_j^{st}(\tau) d\tau, \tag{B17b}$$

Next, let us define $f_j(x,t)dxdt$ as the joint probability that our system is in space interval $(x,\ x+dx)$ and the system undergoes the state transition to state $\Gamma_j$ in time interval $(t,\ t+dt)$. Given that our system is initially located at the origin, $x=0$, in the spatial coordinate, we can write down the following Equations for $f_j(x,t)$ in terms of $\psi_j$, $\psi_j^{st}$, $S_j$ and $S_j^{st}$ [96]

$$f_0(x,t) = \psi_1^{st}(t)\mathcal{G}_1(x,t\,|\,0)p_1 + \int_{-\infty}^{\infty} dx_0 \int_0^t d\tau f_1(x_0,\tau)\psi_1(t-\tau)\mathcal{G}_1(x,t-\tau\,|\,x_0), \tag{B18a}$$

$$f_1(x,t) = \psi_0^{st}(t)\mathcal{G}_0(x,t\,|\,0)p_0 + \int_{-\infty}^{\infty} dx_0 \int_0^t d\tau f_0(x_0,\tau)\psi_0(t-\tau)\mathcal{G}_0(x,t-\tau\,|\,x_0), \tag{B18b}$$

$$f_2(x,t) = 0. \tag{B18c}$$

In Eq. (B18), $\mathcal{G}_j(x,t\,|\,x_0)$ denotes the Green`s function of the diffusion equation with the diffusion constant $D_j$, which satisfies $\partial_t \mathcal{G}_j(x,t\,|\,x_0) = D_j \partial_x^2 \mathcal{G}_j(x,t\,|\,x_0)$ and the initial condition, $\lim_{t\to 0} \mathcal{G}(x,t\,|\,x_0) = \delta(x-x_0)$ . The physical meaning of $\mathcal{G}_j(x,t\,|\,x_0)dx$ is the conditional probability that the system is found in space interval $(x,\ x+dx)$ at time *t*, given that the system is initially located at $x_0$. Note that we have $f_2(x,t)=0$, because there is no transition to state $\Gamma_2$ in our model of the MPM state dynamics, given in Eq. (3).

The joint probability, $P_j(x,t)dx$, that the MPM is at state $\Gamma_j$ and the MPM is located in space interval $(x,\ x+dx)$ at time $t$ is related to $f_j(x,t)dxdt$ as follows:

$$P_j(x,t) = \int_{-\infty}^{\infty} dx_0 \int_0^t d\tau f_j(x_0,\tau) S_j(t-\tau) \mathcal{G}_j(x,t-\tau\,|\,x_0) + S_j^{st}(t) \mathcal{G}_j(x,t,|\,0) p_j\,, \quad \text{(B19)}$$

where $S_j^{st}(t)$ $\left(j \in \{0,1\}\right)$ is given in Eq. (B17) but $S_2^{st}(t)$ are unity at all times.

By taking the Fourier-Laplace transform of Eqs. (B18) and (B19), we obtain

$$\hat{\hat{f}}_0(k,s) = \hat{\psi}_1^{st}(s_1) p_1 + \hat{\hat{f}}_1(k,s) \hat{\psi}_1(s_1)\,, \quad \text{(B20a)}$$

$$\hat{\hat{f}}_1(k,s) = \hat{\psi}_0^{st}(s_0) p_0 + \hat{\hat{f}}_0(k,s) \hat{\psi}_0(s_0)\,, \quad \text{(B20b)}$$

$$\hat{\hat{f}}_2(k,s) = 0\,, \quad \text{(B20c)}$$

$$\hat{\hat{P}}_j(k,s) = \hat{\hat{f}}_j(k,s) \hat{S}_j(s_j) + \hat{S}_j^{st}(s_j) p_j\,, \quad \text{(B21)}$$

with $s_j = s + D_j k^2$. In obtaining Eqs. (B20) and (B21), we have used the convolution theorem and the fact that the Fourier transform of $\mathcal{G}_0(x,t\,|\,0)$ is given by $\tilde{\mathcal{G}}_0(k,t\,|\,0) = e^{-D_0 k^2 t}$ and that the Laplace transform of $h(t)\exp(-at)$ is given by $\hat{h}(s+a)$ for any function $h(t)$. By solving Eqs. (B20) and (B21), we obtain the analytic expression of $\hat{\tilde{P}}(k,s)\left[\equiv \sum_{j=0}^{2} \hat{\tilde{P}}_j(k,s)\right]$ as follows:

$$\hat{\tilde{P}}(k,s) = -\left(1-p_2\right)\frac{\left[1-\hat{\psi}_0(s_0)\right]\left[1-\hat{\psi}_1(s_1)\right]}{\left(\langle t_0\rangle + \langle t_1\rangle\right)\left[1-\hat{\psi}_0(s_0)\hat{\psi}_1(s_1)\right]}\left(\frac{1}{s_0}-\frac{1}{s_1}\right)^2 + \frac{p_0}{s_0} + \frac{p_1}{s_1} + \frac{p_2}{s_2}. \qquad \text{(B22)}$$

For a pedagogical reason, we here present a generalized master equation for $\hat{\tilde{P}}_j(k,s)$ and solve the equation to obtain the analytic expression of $\hat{\tilde{P}}_j(k,s)$. Substituting the definition of $\hat{S}(s) = \left[1-\hat{\psi}(s)\right]/s$ and $\hat{S}^{st}(s) = \left[1-\hat{\psi}^{st}(s)\right]/s$ into Eq. (B21), we obtain

$$\begin{aligned} s_j \hat{\tilde{P}}_j(k,s) - p_j &= \left(1-\hat{\psi}_j(s_j)\right)\hat{\tilde{f}}_j(k,s) - \hat{\psi}_j^{st}(s_j)\,p_j \\ &= \hat{\tilde{f}}_j(k,s) - \hat{\psi}_j(s_j)\hat{\tilde{f}}_j(k,s) - \hat{\psi}_j^{st}(s_j)\,p_j \end{aligned}. \qquad \text{(B23)}$$

Replacing the first term on the right-hand-side (r.h.s.) of Eq. (B23) with the r.h.s. of Eq. (B20) we obtain

$$s_0\hat{\tilde{P}}_0(k,s) - p_0 = \hat{\psi}_1(s_1)\hat{\tilde{f}}_1(k,s) - \hat{\psi}_0(s_0)\hat{\tilde{f}}_0(k,s) + \hat{\psi}_1^{st}(s_1)\,p_1 - \hat{\psi}_0^{st}(s_0)\,p_0\,, \qquad \text{(B24a)}$$

$$s_1\hat{\tilde{P}}_1(k,s) - p_1 = \hat{\psi}_0(s_0)\hat{\tilde{f}}_0(k,s) - \hat{\psi}_1(s_1)\hat{\tilde{f}}_1(k,s) + \hat{\psi}_0^{st}(s_0)\,p_0 - \hat{\psi}_1^{st}(s_1)\,p_1\,. \qquad \text{(B24b)}$$

$$s_2\hat{\tilde{P}}_2(k,s) - p_2 = 0 \qquad \text{(B24c)}$$

Using the first equality of Eq. (B23), we can represent $\hat{\hat{f}}_0(k,s)$ and $\hat{\hat{f}}_1(k,s)$ in terms of $\hat{\hat{P}}_0(k,s)$ and $\hat{\hat{P}}_1(k,s)$, i.e.,

$$\hat{\hat{f}}_j(k,s) = \frac{s_j}{1-\hat{\psi}_j(s_j)} \hat{\hat{P}}_j(k,s) - \frac{1-\hat{\psi}_j^{st}(s_j)}{1-\hat{\psi}_j(s_j)} p_j . \tag{B25}$$

Substituting Eq. (B25) into the r.h.s. of Eq. (B24), we obtain

$$\begin{aligned} s_0\hat{\hat{P}}_0(k,s) - p_0 = {} & \frac{s_1\hat{\psi}_1(s_1)}{1-\hat{\psi}_1(s_1)} \hat{\hat{P}}_1(k,s) - \frac{s_0\hat{\psi}_0(s_0)}{1-\hat{\psi}_0(s_0)} \hat{\hat{P}}_0(k,s) \\ & + \frac{1}{s_1}\left[\frac{1}{\langle t_1\rangle} - \frac{s_1\hat{\psi}_1(s_1)}{1-\hat{\psi}_1(s_1)}\right] p_1 - \frac{1}{s_0}\left[\frac{1}{\langle t_0\rangle} - \frac{s_0\hat{\psi}_0(s_0)}{1-\hat{\psi}_0(s_0)}\right] p_0 , \end{aligned} \tag{B26a}$$

$$\begin{aligned} s_1\hat{\hat{P}}_1(k,s) - p_1 = {} & \frac{s_0\hat{\psi}_0(s_0)}{1-\hat{\psi}_0(s_0)} \hat{\hat{P}}_0(k,s) - \frac{s_1\hat{\psi}_1(s_1)}{1-\hat{\psi}_1(s_1)} \hat{\hat{P}}_1(k,s) \\ & + \frac{1}{s_0}\left[\frac{1}{\langle t_0\rangle} - \frac{s_0\hat{\psi}_0(s_0)}{1-\hat{\psi}_0(s_0)}\right] p_0 - \frac{1}{s_1}\left[\frac{1}{\langle t_1\rangle} - \frac{s_1\hat{\psi}_1(s_1)}{1-\hat{\psi}_1(s_1)}\right] p_1 . \end{aligned} \tag{B26b}$$

Finally, replacing $s_j$ with $s + D_j k^2$ on the left-hand-side of Eqs. (B26) and (B24c), we obtain the following equations for $\{\hat{\hat{P}}_0, \hat{\hat{P}}_1, \hat{\hat{P}}_2\}$

$$\begin{aligned} s\hat{\hat{P}}_0(k,s) - p_0 = {} & -D_0k^2\hat{\hat{P}}_0(k,s) + \hat{\hat{K}}_1(k,s)\hat{\hat{P}}_1(k,s) - \hat{\hat{K}}_0(k,s)\hat{\hat{P}}_0(k,s) \\ & + \hat{\hat{T}}_1(k,s)p_1 - \hat{\hat{T}}_0(k,s)p_0 \end{aligned} , \tag{B27a}$$

$$\begin{aligned} s\hat{\hat{P}}_1(k,s) - p_1 = {} & -D_1k^2\hat{\hat{P}}_1(k,s) + \hat{\hat{K}}_0(k,s)\hat{\hat{P}}_0(k,s) - \hat{\hat{K}}_1(k,s)\hat{\hat{P}}_1(k,s) \\ & + \hat{\hat{T}}_0(k,s)p_0 - \hat{\hat{T}}_1(k,s)p_1 \end{aligned} , \tag{B27b}$$

$$s\hat{\hat{P}}_2(k,s) - p_2 = -D_2k^2\hat{\hat{P}}_2(k,s) . \tag{B27c}$$

with $\hat{\tilde{K}}_i(k,s) = s_i\hat{\psi}_i(s_i)\big/\left[1-\hat{\psi}_i(s_i)\right]$ and $\hat{\tilde{T}}_i(k,s) = s_i^{-1}\left(\langle t_i\rangle^{-1} - s_i\hat{\psi}_i(s_i)\big/\left[1-\hat{\psi}_i(s_i)\right]\right)$.

Noting that the l.h.s. of Eq. (B27) is the Laplace transform, $\hat{\dot{\tilde{P}}}_j$, of $\partial_t\tilde{P}_j(k,t)$, we obtain the following generalized master equation

$$\begin{aligned}\hat{\dot{\tilde{\mathbf{P}}}}(k,s) &= s\hat{\tilde{\mathbf{P}}}(k,s)-\mathbf{p}\\ &= \left[-k^2\mathbf{D}+\hat{\tilde{\mathbf{K}}}(k,s)\right]\cdot\hat{\tilde{\mathbf{P}}}(k,s)+\hat{\tilde{\mathbf{T}}}(k,s)\cdot\mathbf{p}\end{aligned}, \tag{B28}$$

where the vector, $\hat{\tilde{\mathbf{P}}}(k,s)$, $\mathbf{p}$ and the matrices, $\mathbf{D}$, $\hat{\tilde{\mathbf{K}}}(k,s)$ and $\hat{\tilde{\mathbf{T}}}(k,s)$ are given by

$$\hat{\tilde{\mathbf{P}}}(k,s) \equiv \begin{pmatrix}\hat{\tilde{P}}_0(k,s)\\ \hat{\tilde{P}}_1(k,s)\\ \hat{\tilde{P}}_2(k,s)\end{pmatrix}, \tag{B29a}$$

$$\mathbf{p} \equiv \begin{pmatrix}p_0\\ p_1\\ p_2\end{pmatrix}, \tag{B29b}$$

$$\mathbf{D} \equiv \begin{pmatrix}D_0 & 0 & 0\\ 0 & D_1 & 0\\ 0 & 0 & D_2\end{pmatrix}, \tag{B29c}$$

$$\hat{\tilde{\mathbf{K}}}(k,s) \equiv \begin{pmatrix}-\hat{\tilde{K}}_0(k,s) & \hat{\tilde{K}}_1(k,s) & 0\\ \hat{\tilde{K}}_0(k,s) & -\hat{\tilde{K}}_1(k,s) & 0\\ 0 & 0 & 0\end{pmatrix}, \tag{B29d}$$

and

$$\hat{\tilde{\mathbf{T}}}(k,s) \equiv \begin{pmatrix} -\hat{\tilde{T}}_0(k,s) & \hat{\tilde{T}}_1(k,s) & 0 \\ \hat{\tilde{T}}_0(k,s) & -\hat{\tilde{T}}_1(k,s) & 0 \\ 0 & 0 & 0 \end{pmatrix}. \tag{B29e}$$

We obtain the following expression of $\hat{\tilde{\mathbf{P}}}(k,s)$ from Eq. (B28)

$$\hat{\tilde{\mathbf{P}}}(k,s) = \left[ s\mathbf{I} + k^2\mathbf{D} - \hat{\tilde{\mathbf{K}}}(k,s) \right]^{-1} \cdot \left[ \mathbf{I} + \hat{\tilde{\mathbf{T}}}(k,s) \right] \cdot \mathbf{p}^i . \tag{B30}$$

From Eq. (B30), we recover the analytic expression of $\hat{\tilde{P}}(k,s) \left[ \equiv \sum_{i=1}^{3} \hat{\tilde{P}}_i(k,s) \right]$ given in Eq. (B22).

**10. Analytic expression of the vesicle displacement distribution in hyperphosphorylated cells**

Given that the position of a vesicle can be represented by $d_x = R_x + x'$ where $R_x$ and $x'$ denote, respectively, the position of the MPM and the relative position of the vesicle with respect to the MPM, the displacement distribution $g(d_x,t)$ of the vesicle-MPM along the microtubule can be written

$$\begin{aligned} g(d_x,t) &= \int_{-\infty}^{\infty} dR_x \int_{-\infty}^{\infty} dx' \delta(d_x - R_x - x') P(R_x,t) f(x') \\ &= \int_{-\infty}^{\infty} dx' P(d_x - x', t) f(x') \end{aligned}, \tag{B31}$$

where $\delta(x)$, $P(R_x,t)$ and $f(x')$ denote, respectively, Dirac's delta function, the distribution of the MPM displacement, and the distribution of the relative position $x'$, of the vesicle with respect to the MPM. Taking the Fourier-Laplace transform of Eq. (B31), we obtain

$$\hat{\tilde{g}}(k,s) = \hat{\tilde{P}}(k,s) \tilde{f}(k), \tag{B32}$$

This equation with Eqs. (B14) and (B15) provides the analytic expression of the displacement distribution of the vesicle transported by MPM in the Fourier-Laplace domain.

We confirm the correctness of Eq. (B32) against an accurate stochastic simulation result for our model (Fig. 12). Numerical inversion[97] of Eq. (B32) using the optimized parameter values in Table 1 allows us to predict the spatial profile of the vesicle displacement distribution. As shown in Fig. 2(d), our theoretical prediction is in excellent agreement with the experimental results for the vesicle displacement distribution.

**11. Ballistic motion of the random walker**

Let us consider a random walker moving on a one-dimensional lattice with lattice spacing $\varepsilon$ at a given environmental state $\Gamma$. The random walker can jump to one of two nearest neighboring sites. Let $\varphi_{n,\Gamma}^{L(R)}(t)dt$ denote the probability that the random walker jumps from the *n*th site to the $(n-1)$th site [the $(n+1)$th site] between time $t \sim t+dt$ in the absence of jump to the $(n+1)$th site [the $(n-1)$th site] at a given environmental state $\Gamma$. When the random walker can jump in both directions, the corresponding waiting time distribution of the jump is then given by $\Phi_{n,\Gamma}^{L(R)}(t) = \varphi_{n,\Gamma}^{L(R)}(t)\int_t^\infty \varphi_{n,\Gamma}^{R(L)}(\tau)\,d\tau$. $\Phi_{n,\Gamma}(t)[\equiv \Phi_{n,\Gamma}^{L}(t) + \Phi_{n,\Gamma}^{R}(t)]$ is nothing but the normalized waiting time distribution for a random walker's jump at the *n*th site. In spatially homogeneous systems, $\Phi_{n,\Gamma}(t)$ is independent of *n*, i.e., $\Phi_{n,\Gamma}(t) = \Phi_{\Gamma}(t)$.

To derive the transport equation associated with the motion of a random walker, let us consider the probability, $f_\Gamma(n,t\,|\,n_0)dt$, that a random walker, initially located at the $n_0$th site, arrives at the $n$th site between time $t \sim t+dt$ at a given environmental state $\Gamma$. $f_\Gamma(n,t|n_0)$ satisfies the following equation[98]:

$$\hat{f}_\Gamma(n,s|n_0) = \hat{\Phi}_\Gamma(s)\Big(R\hat{f}_\Gamma(n-1,s|n_0) + L\hat{f}_\Gamma(n+1,s|n_0)\Big) + \delta_{n,n_0}, \quad \text{(B33)}$$

where $R$ and $L$ respectively designate the probability of a random walker's jump to the right and left directions; in our case, $R = L = 1/2$. In equation (B33), $\hat{f}_\Gamma(s)$ denotes the Laplace transform of $f_\Gamma(t)$, i.e., $\hat{f}_\Gamma(s) = \int_0^\infty e^{-st} f_\Gamma(t)dt$. $f_\Gamma(n,t|n_0)$ is related to the probability, $g_\Gamma(n,t|n_0)$, that a random walker, initially located at the $n_0$th site, is found at the $n$th site at time $t$ at a given environmental state $\Gamma$ by $\hat{g}_\Gamma(n,s|n_0) = [1-\hat{\Phi}_\Gamma(s)]\hat{f}_\Gamma(n,s|n_0)/s$ in the Laplace domain. Here, $[1-\hat{\Phi}_\Gamma(s)]/s$ is the Laplace transform of $1-\int_0^t \Phi_\Gamma(\tau)d\tau$, which means the probability that a random walker does not jump by elapsed time $t$ since it arrives at a lattice site. Substituting $\hat{f}_\Gamma(n,s|n_0) = s\hat{g}_\Gamma(n,s|n_0)/[1-\hat{\Phi}_\Gamma(s)]$ into Eq. (B33), we can obtain the following generalized master equation:

$$\hat{\dot{g}}_\Gamma(n,s|n_0) = \frac{1}{2}\frac{s\hat{\Phi}_\Gamma(s)}{1-\hat{\Phi}_\Gamma(s)}\big(\hat{g}_\Gamma(n-1,s|n_0) + \hat{g}_\Gamma(n+1,s|n_0) - 2\hat{g}_\Gamma(n,s|n_0)\big), \quad \text{(B34)}$$

where $\hat{\dot{g}}_\Gamma(n,s|n_0)$ denotes the Laplace transform of the first-order time derivative of $g_\Gamma(n,t|n_0)$, i.e., $\hat{\dot{g}}_\Gamma(n,s|n_0) = s\hat{g}_\Gamma(n,s|n_0) - \delta_{n,n_0}$. In the continuum limit, Eq. (B34) can be expressed as

$$\hat{\dot{p}}(x,\mathbf{\Gamma},s) = \hat{\mathcal{D}}_\Gamma(s)\frac{\partial^2}{\partial x^2}\hat{p}(x,\mathbf{\Gamma},s), \quad \text{(B35)}$$

where $\hat{\mathcal{D}}_{\mathbf{\Gamma}}(s)$ denotes the diffusion kernel defined by $\hat{\mathcal{D}}_{\Gamma}(s) = (\varepsilon^2/2)s\hat{\Phi}_{\Gamma}(s)/[1-\hat{\Phi}_{\Gamma}(s)]$, and $\hat{p}(x,\mathbf{\Gamma},s)$ represents the probability density function of the random walker at a given environmental state $\Gamma$, which has $p(x,\mathbf{\Gamma},0) = \delta(x-x_0)$ as initial condition, with $x = \varepsilon n$, $x_0 = \varepsilon n_0$, and $g_{\mathbf{\Gamma}}(n,t\,|\,n_0) = \varepsilon p(x,\mathbf{\Gamma},t)$.

If a random walker exhibits ballistic motion, every jump event occurs at a regular interval, specifically $\tau_{\mathbf{\Gamma}}$, i.e., $\Phi_{\mathbf{\Gamma}}(t) = \delta(t-\tau_{\mathbf{\Gamma}})$. In this case, the diffusion kernel follows:

$$\hat{\mathcal{D}}_{\mathbf{\Gamma}}(s) = \frac{\varepsilon^2}{2}\frac{se^{-s\tau_{\Gamma}}}{1-e^{-s\tau_{\Gamma}}}. \tag{B36}$$

At time scales far longer than $\tau_{\Gamma}$, Eq. (B35) reduces to

$$\dot{\hat{p}}(x,\mathbf{\Gamma},s) = D_{\mathbf{\Gamma}}\frac{\partial^2}{\partial x^2}\hat{p}(x,\mathbf{\Gamma},s), \tag{B37}$$

where $D_{\mathbf{\Gamma}}$ denotes the $\mathbf{\Gamma}$-dependent diffusion coefficient given by the small-*s* limiting value of the diffusion kernel, $\hat{\mathcal{D}}_{\mathbf{\Gamma}}(s)$, explicitly, $D_{\mathbf{\Gamma}} = \hat{\mathcal{D}}_{\mathbf{\Gamma}}(0) = \varepsilon^2/2\tau_{\mathbf{\Gamma}}$. When an environmental state coupled to the motion of a random walker varies over time, Eq. (B37) can be cast into the following form[62]:

$$\dot{\hat{p}}(x,\mathbf{\Gamma},s) = D_{\mathbf{\Gamma}}\frac{\partial^2}{\partial x^2}\hat{p}(x,\mathbf{\Gamma},s) + L(\mathbf{\Gamma})\hat{p}(x,\mathbf{\Gamma},s), \tag{B38}$$

where $L(\mathbf{\Gamma})$ denotes the time evolution operator governing dynamics of environmental state $\mathbf{\Gamma}$.

**12. Simulation method for vesicle displacement distribution**

Here, we present the stochastic simulation algorithm to obtain the displacement distribution of vesicles carried by the motor protein multiplex in forskolin-treated cells using our optimized model. Let us first consider the MPM motion along the microtubule. The initial state of the MPM-microtubule system is sampled according to the steady-state probability, $p_i$, of MPM state $\Gamma_i$, given in Table 1. The initial position, $R_x(t=0)$, of the MPM along the microtubule is set to be zero. While our MPM-microtubule system is at state $\Gamma_j$, the MPM undergoes Brownian motion along the microtubule with diffusion coefficient $D_j$. The lifetimes of states $\Gamma_0$ and $\Gamma_1$ are sampled from gamma distributions with parameter values given in Table 1. The lifetime of state $\Gamma_2$ is set to be infinite. During the lifetime time of state $\Gamma_j$, the time trajectories of the MPM position are generated using the standard simulation method of over-damped Brownian motion, i.e.,

$$R_x(t+\Delta t) = R_x(t) + \sqrt{2D_j \Delta t} N(0,1), \tag{B39}$$

where $\Delta t$ and $N(0,1)$ denote, respectively, the time step size of the simulation and the Gaussian random number with zero mean and unit variance. In our simulation, we set $\Delta t = 0.01\,\text{s}$. Upon MPM state transition from $\Gamma_i$ to $\Gamma_j$, we change the diffusion coefficient from $D_i$ to $D_j$, and perform the Brownian dynamics simulation with diffusion coefficient $D_j$ as of the sampled lifetime of $\Gamma_j$.

Because the vesicle undergoes fast liberational motion around the MPM that it is attached, the displacement, $d_x$, of a vesicle can be represented by $d_x = R_x + x'$, where $R_x$ and $x'$ denote the displacement of the motor protein multiplex (MPM) and the change in the relative position of the vesicle with respect to the MPM, respectively. In our simulation, the distance $x'$ between the MPM and the vesicle is sampled according to probability distribution $f(x')$ given in Eq. (B14), i.e.,

$$f(x') = \int_{-\infty}^{\infty} dq \sqrt{\frac{\kappa_0}{2\pi(1+q^2)}} \exp\left(-\frac{\kappa_0 x'^2}{2(1+q^2)}\right) N(q \mid \sigma_q^2), \tag{B14}$$

with $\kappa_0^{-1} = 2.4\times10^{-3}\mu\text{m}^2$ and $\sigma_q^2 = 2.46$. For this purpose, we first calculate the value of $\kappa$ defined by $\kappa \equiv \kappa_0 / [1+\sigma_q^2 N(0,1)^2]$, and sample the value of $x'$ by $x' = \kappa^{-1/2} N(0,1)$. To get the position of the vesicle, $d_x$, we add up $x'$ and $R_x$ in Eq. (B39), i.e., $d_x = R_x + x'$. We repeatedly perform the stochastic simulation to generate a million of trajectories. As shown in Fig. 12, our analytic results given in Eq. (B32) is in perfect agreement with this stochastic simulation results for the vesicle displacement distribution of our optimized model.

**References**

[1] N. Hirokawa, Y. Noda, Y. Tanaka, and S. Niwa, *Kinesin superfamily motor proteins and intracellular transport*, Nat. Rev. Mol. Cell Biol. **10**, 682 (2009).
[2] C. L. Asbury, A. N. Fehr, and S. M. Block, *Kinesin moves by an asymmetric hand-over-hand mechanism*, Science **302**, 2130 (2003).
[3] M. Vershinin, B. C. Carter, D. S. Razafsky, S. J. King, and S. P. Gross, *Multiple-motor based transport and its regulation by Tau*, Proc. Natl. Acad. Sci. U.S.A. **104**, 87 (2007).
[4] L. Kaplan, A. Ierokomos, P. Chowdary, Z. Bryant, and B. Cui, *Rotation of endosomes demonstrates coordination of molecular motors during axonal transport*, Sci. Adv. **4**, e1602170 (2018).
[5] B. H. Blehm and P. R. Selvin, *Single-molecule fluorescence and in vivo optical traps: how multiple dyneins and kinesins interact*, Chem. Rev. **114**, 3335 (2014).
[6] K. I. Schimert, B. G. Budaitis, D. N. Reinemann, M. J. Lang, and K. J. Verhey, *Intracellular cargo transport by single-headed kinesin motors*, Proc. Natl. Acad. Sci. U.S.A. **116**, 6152 (2019).
[7] R. D. Vale, *The molecular motor toolbox for intracellular transport*, Cell **112**, 467 (2003).
[8] P. W. Baas and S. Lin, *Hooks and comets: The story of microtubule polarity orientation in the neuron*, Dev Neurobiol **71**, 403 (2011).
[9] P. W. Baas, J. S. Deitch, M. M. Black, and G. A. Banker, *Polarity orientation of microtubules in hippocampal neurons: uniformity in the axon and nonuniformity in the dendrite*, Proc. Natl. Acad. Sci. U.S.A. **85**, 8335 (1988).
[10] D. Chrétien, J. M. Kenney, S. D. Fuller, and R. H. Wade, *Determination of microtubule polarity by cryo-electron microscopy*, Structure **4**, 1031 (1996).
[11] M. J. Shelley, *The Dynamics of Microtubule/Motor-Protein Assemblies in Biology and Physics*, Annu. Rev. Fluid Mech. **48**, 487 (2016).
[12] S. Fürthauer, B. Lemma, P. J. Foster, S. C. Ems-McClung, C.-H. Yu, C. E. Walczak, Z. Dogic, D. J. Needleman, and M. J. Shelley, *Self-straining of actively crosslinked microtubule networks*, Nat. Phys. **15**, 1295 (2019).
[13] X. Nan, P. A. Sims, and X. S. Xie, *Organelle tracking in a living cell with microsecond time resolution and nanometer spatial precision*, ChemPhysChem **9**, 707 (2008).
[14] P. A. Sims and X. S. Xie, *Probing dynein and kinesin stepping with mechanical manipulation in a living cell*, ChemPhysChem **10**, 1511 (2009).
[15] S. Klumpp and R. Lipowsky, *Cooperative cargo transport by several molecular motors*, Proc. Natl. Acad. Sci. U.S.A. **102**, 17284 (2005).
[16] M. J. Müller, S. Klumpp, and R. Lipowsky, *Tug-of-war as a cooperative mechanism for bidirectional cargo transport by molecular motors*, Proc. Natl. Acad. Sci. U.S.A. **105**, 4609 (2008).
[17] K. Shin, S. Song, Y. H. Song, S. Hahn, J.-H. Kim, G. Lee, I.-C. Jeong, J. Sung, and K. T. Lee, *Anomalous dynamics of in vivo cargo delivery by motor protein multiplexes*, J. Phys. Chem, Lett. **10**, 3071 (2019).
[18] N. Hirokawa, *Kinesin and dynein superfamily proteins and the mechanism of organelle transport*, Science **279**, 519 (1998).

[19] C. Kural, H. Kim, S. Syed, G. Goshima, V. I. Gelfand, and P. R. Selvin, *Kinesin and dynein move a peroxisome in vivo: a tug-of-war or coordinated movement?*, Science **308**, 1469 (2005).
[20] W. O. Hancock, *Bidirectional cargo transport: moving beyond tug of war*, Nat. Rev. Mol. Cell Biol. **15**, 615 (2014).
[21] K. Chen, B. Wang, and S. Granick, *Memoryless self-reinforcing directionality in endosomal active transport within living cells*, Nat. Mater. **14**, 589 (2015).
[22] X. Nan, P. A. Sims, P. Chen, and X. S. Xie, *Observation of individual microtubule motor steps in living cells with endocytosed quantum dots*, J. Phys. Chem. B **109**, 24220 (2005).
[23] B. Cui, C. Wu, L. Chen, A. Ramirez, E. L. Bearer, W.-P. Li, W. C. Mobley, and S. Chu, *One at a time, live tracking of NGF axonal transport using quantum dots*, Proc. Natl. Acad. Sci. U.S.A. **104**, 13666 (2007).
[24] P. D. Chowdary, D. L. Che, L. Kaplan, O. Chen, K. Pu, M. Bawendi, and B. Cui, *Nanoparticle-assisted optical tethering of endosomes reveals the cooperative function of dyneins in retrograde axonal transport*, Sci. Rep. **5**, 1 (2015).
[25] M. S. Song, H. C. Moon, J.-H. Jeon, and H. Y. Park, *Neuronal messenger ribonucleoprotein transport follows an aging Lévy walk*, Nat. Commun. **9**, 1 (2018).
[26] S. J. Park, S. Song, I.-C. Jeong, H. R. Koh, J.-H. Kim, and J. Sung, *Nonclassical kinetics of clonal yet heterogeneous enzymes*, J. Phys. Chem, Lett. **8**, 3152 (2017).
[27] J. Avila, J. J. Lucas, M. Perez, and F. Hernandez, *Role of tau protein in both physiological and pathological conditions*, Physiological reviews (2004).
[28] S. Khatoon, I. Grundke-Iqbal, and K. Iqbal, *Brain levels of microtubule-associated protein τ are elevated in Alzheimer's disease: A radioimmuno-slot-blot assay for nanograms of the protein*, Journal of neurochemistry **59**, 750 (1992).
[29] S. Khatoon, I. Grundke-Iqbal, and K. Iqbal, *Levels of normal and abnormally phosphorylated tau in different cellular and regional compartments of Alzheimer disease and control brains*, FEBS Lett. **351**, 80 (1994).
[30] T. Arendt, J. T. Stieler, and M. Holzer, *Tau and tauopathies*, Brain research bulletin **126**, 238 (2016).
[31] M. Goedert, M. G. Spillantini, R. Jakes, D. Rutherford, and R. Crowther, *Multiple isoforms of human microtubule-associated protein tau: sequences and localization in neurofibrillary tangles of Alzheimer's disease*, Neuron **3**, 519 (1989).
[32] A. d. C. Alonso, T. Zaidi, M. Novak, I. Grundke-Iqbal, and K. Iqbal, *Hyperphosphorylation induces self-assembly of τ into tangles of paired helical filaments/straight filaments*, Proc. Natl. Acad. Sci. U.S.A. **98**, 6923 (2001).
[33] H. Braak, I. Alafuzoff, T. Arzberger, H. Kretzschmar, and K. Del Tredici, *Staging of Alzheimer disease-associated neurofibrillary pathology using paraffin sections and immunocytochemistry*, Acta neuropathologica **112**, 389 (2006).
[34] A. De Calignon *et al.*, *Propagation of tau pathology in a model of early Alzheimer's disease*, Neuron **73**, 685 (2012).
[35] Y. Wang and E. Mandelkow, *Tau in physiology and pathology*, Nat. Rev. Neurosci. **17**, 22 (2016).
[36] S. Lesné, M. T. Koh, L. Kotilinek, R. Kayed, C. G. Glabe, A. Yang, M. Gallagher, and K. H. Ashe, *A specific amyloid-β protein assembly in the brain impairs memory*, Nature **440**, 352 (2006).

[37] C. Piller, *Blots on a field?*, Science (New York, NY) **377**, 358 (2022).
[38] J.-z. Wang, Q. Wu, A. Smith, I. Grundke-Iqbal, and K. Iqbal, *τ is phosphorylated by GSK-3 at several sites found in Alzheimer disease and its biological activity markedly inhibited only after it is prephosphorylated by A-kinase*, FEBS Lett. **436**, 28 (1998).
[39] B. Falcon *et al.*, *Conformation determines the seeding potencies of native and recombinant Tau aggregates*, J. Biol. Chem. **290**, 1049 (2015).
[40] T. F. Gendron and L. Petrucelli, *The role of tau in neurodegeneration*, Mol. Neurodegener. **4**, 1 (2009).
[41] A. L. Houck, F. Hernández, and J. Ávila, *A simple model to study tau pathology*, Journal of experimental neuroscience **10**, JEN. S25100 (2016).
[42] S.-W. Min *et al.*, *Critical role of acetylation in tau-mediated neurodegeneration and cognitive deficits*, Nature medicine **21**, 1154 (2015).
[43] D. J. Irwin, *Tauopathies as clinicopathological entities*, Parkinsonism & related disorders **22**, S29 (2016).
[44] L. I. Binder, A. L. Guillozet-Bongaarts, F. Garcia-Sierra, and R. W. Berry, *Tau, tangles, and Alzheimer's disease*, Biochim. Biophys. Acta Mol. Basis Dis. **1739**, 216 (2005).
[45] K. J. De Vos, A. J. Grierson, S. Ackerley, and C. C. Miller, *Role of axonal transport in neurodegenerative diseases*, Annu. Rev. Neurosci. **31**, 151 (2008).
[46] S. Maday, A. E. Twelvetrees, A. J. Moughamian, and E. L. Holzbaur, *Axonal transport: cargo-specific mechanisms of motility and regulation*, Neuron **84**, 292 (2014).
[47] A. Ebneth, R. Godemann, K. Stamer, S. Illenberger, B. Trinczek, E.-M. Mandelkow, and E. Mandelkow, *Overexpression of tau protein inhibits kinesin-dependent trafficking of vesicles, mitochondria, and endoplasmic reticulum: implications for Alzheimer's disease*, J. Cell Biol. **143**, 777 (1998).
[48] M. Vershinin, J. Xu, D. S. Razafsky, S. J. King, and S. P. Gross, *Tuning microtubule-based transport through filamentous MAPs: the problem of dynein*, Traffic **9**, 882 (2008).
[49] A. R. Chaudhary, F. Berger, C. L. Berger, and A. G. Hendricks, *Tau directs intracellular trafficking by regulating the forces exerted by kinesin and dynein teams*, Traffic **19**, 111 (2018).
[50] R. Dixit, J. L. Ross, Y. E. Goldman, and E. L. Holzbaur, *Differential regulation of dynein and kinesin motor proteins by tau*, Science **319**, 1086 (2008).
[51] S. Ahn, J.-S. Suh, Y.-K. Jang, H. Kim, K. Han, Y. Lee, G. Choi, and T.-J. Kim, *TAUCON and TAUCOM: A novel biosensor based on fluorescence resonance energy transfer for detecting tau hyperphosphorylation-associated cellular pathologies*, Biosensors and Bioelectronics **237**, 115533 (2023).
[52] E. Barkai, Y. Garini, and R. Metzler, *Strange kinetics of single molecules in living cells*, Physics today **65**, 29 (2012).
[53] I. Goychuk, V. O. Kharchenko, and R. Metzler, *Molecular motors pulling cargos in the viscoelastic cytosol: how power strokes beat subdiffusion*, Physical Chemistry Chemical Physics **16**, 16524 (2014).
[54] H. Park, E. Toprak, and P. R. Selvin, *Single-molecule fluorescence to study molecular motors*, Quarterly reviews of biophysics **40**, 87 (2007).
[55] S. W. Hell and J. Wichmann, *Breaking the diffraction resolution limit by stimulated emission: stimulated-emission-depletion fluorescence microscopy*, Opt. Lett. **19**, 780 (1994).

[56] E. Betzig and R. J. Chichester, *Single molecules observed by near-field scanning optical microscopy*, Science **262**, 1422 (1993).
[57] M. J. Rust, M. Bates, and X. Zhuang, *Sub-diffraction-limit imaging by stochastic optical reconstruction microscopy (STORM)*, Nat. Methods **3**, 793 (2006).
[58] A. Yildiz, J. N. Forkey, S. A. McKinney, T. Ha, Y. E. Goldman, and P. R. Selvin, *Myosin V walks hand-over-hand: single fluorophore imaging with 1.5-nm localization*, science **300**, 2061 (2003).
[59] M. P. Gordon, T. Ha, and P. R. Selvin, *Single-molecule high-resolution imaging with photobleaching*, Proc. Natl. Acad. Sci. U.S.A. **101**, 6462 (2004).
[60] B. P. English, W. Min, A. M. Van Oijen, K. T. Lee, G. Luo, H. Sun, B. J. Cherayil, S. Kou, and X. S. Xie, *Ever-fluctuating single enzyme molecules: Michaelis-Menten equation revisited*, Nat. Chem. Biol **2**, 87 (2006).
[61] X. Mao, C. Liu, M. Hesari, N. Zou, and P. Chen, *Super-resolution imaging of non-fluorescent reactions via competition*, Nat. Chem. **11**, 687 (2019).
[62] S. Song, S. J. Park, M. Kim, J. S. Kim, B. J. Sung, S. Lee, J.-H. Kim, and J. Sung, *Transport dynamics of complex fluids*, Proc. Natl. Acad. Sci. U.S.A. **116**, 12733 (2019).
[63] B. Wang, S. M. Anthony, S. C. Bae, and S. Granick, *Anomalous yet Brownian*, Proc. Natl. Acad. Sci. U.S.A. **106**, 15160 (2009).
[64] J. Guan, B. Wang, and S. Granick, *Even hard-sphere colloidal suspensions display Fickian yet non-Gaussian diffusion*, ACS nano **8**, 3331 (2014).
[65] N. Shemesh, T. Adiri, and Y. Cohen, *Probing microscopic architecture of opaque heterogeneous systems using double-pulsed-field-gradient NMR*, J. Am. Chem. Soc. **133**, 6028 (2011).
[66] M. Shayegan, R. Tahvildari, K. Metera, L. Kisley, S. W. Michnick, and S. R. Leslie, *Probing inhomogeneous diffusion in the microenvironments of phase-separated polymers under confinement*, J. Am. Chem. Soc. **141**, 7751 (2019).
[67] R. Pastore, A. Ciarlo, G. Pesce, F. Greco, and A. Sasso, *Rapid Fickian yet non-Gaussian diffusion after subdiffusion*, Phys. Rev. Lett. **126**, 158003 (2021).
[68] A. V. Chechkin, F. Seno, R. Metzler, and I. M. Sokolov, *Brownian yet non-Gaussian diffusion: from superstatistics to subordination of diffusing diffusivities*, Physical Review X **7**, 021002 (2017).
[69] C. Xue, X. Zheng, K. Chen, Y. Tian, and G. Hu, *Probing non-Gaussianity in confined diffusion of nanoparticles*, J. Phys. Chem, Lett. **7**, 514 (2016).
[70] G. Kwon, B. J. Sung, and A. Yethiraj, *Dynamics in crowded environments: is non-Gaussian Brownian diffusion normal?*, J. Phys. Chem. B **118**, 8128 (2014).
[71] B. Wang, J. Kuo, S. C. Bae, and S. Granick, *When Brownian diffusion is not Gaussian*, Nat. Mater. **11**, 481 (2012).
[72] I. Goychuk and P. Hänggi, *Non-Markovian stochastic resonance*, Phys. Rev. Lett. **91**, 070601 (2003).
[73] A. d. C. Alonso, T. Zaidi, I. Grundke-Iqbal, and K. Iqbal, *Role of abnormally phosphorylated tau in the breakdown of microtubules in Alzheimer disease*, Proc. Natl. Acad. Sci. U.S.A. **91**, 5562 (1994).
[74] J. Gilley *et al.*, *Mislocalization of neuronal tau in the absence of tangle pathology in phosphomutant tau knockin mice*, Neurobiology of aging **39**, 1 (2016).

[75] M. E. Orr, A. C. Sullivan, and B. Frost, *A brief overview of tauopathy: causes, consequences, and therapeutic strategies*, Trends Pharmacol. Sci. **38**, 637 (2017).
[76] J. S. Rothman, L. Kocsis, E. Herzog, Z. Nusser, and R. A. Silver, *Physical determinants of vesicle mobility and supply at a central synapse*, Elife **5**, e15133 (2016).
[77] S. J. Park, S. Song, G.-S. Yang, P. M. Kim, S. Yoon, J.-H. Kim, and J. Sung, *The chemical fluctuation theorem governing gene expression*, Nat. Commun. **9**, 1 (2018).
[78] G. Rodriguez, J. A. Ross, Z. S. Nagy, and R. A. Kirken, *Forskolin-inducible cAMP pathway negatively regulates T-cell proliferation by uncoupling the interleukin-2 receptor complex*, J. Biol. Chem. **288**, 7137 (2013).
[79] J. Vandamme, D. Castermans, and J. M. Thevelein, *Molecular mechanisms of feedback inhibition of protein kinase A on intracellular cAMP accumulation*, Cellular signalling **24**, 1610 (2012).
[80] D. P. Hanger, J. C. Betts, T. L. Loviny, W. P. Blackstock, and B. H. Anderton, *New phosphorylation sites identified in hyperphosphorylated tau (paired helical filament-tau) from Alzheimer's disease brain using nanoelectrospray mass spectrometry*, Journal of neurochemistry **71**, 2465 (1998).
[81] D. P. Hanger, H. L. Byers, S. Wray, K.-Y. Leung, M. J. Saxton, A. Seereeram, C. H. Reynolds, M. A. Ward, and B. H. Anderton, *Novel phosphorylation sites in tau from Alzheimer brain support a role for casein kinase 1 in disease pathogenesis*, J. Biol. Chem. **282**, 23645 (2007).
[82] M. Hasegawa, M. Morishima-Kawashima, K. Takio, M. Suzuki, K. a. Titani, and Y. Ihara, *Protein sequence and mass spectrometric analyses of tau in the Alzheimer's disease brain*, J. Biol. Chem. **267**, 17047 (1992).
[83] M. Morishima-Kawashima, M. Hasegawa, K. Takio, M. Suzuki, H. Yoshida, K. Titani, and Y. Ihara, *Proline-directed and Non-proline-directed Phosphorylation of PHF-tau (∗)*, J. Biol. Chem. **270**, 823 (1995).
[84] J. Z. Wang, I. Grundke-Iqbal, and K. Iqbal, *Kinases and phosphatases and tau sites involved in Alzheimer neurofibrillary degeneration*, European Journal of Neuroscience **25**, 59 (2007).
[85] Q.-G. Ren, X.-M. Liao, X.-Q. Chen, G.-P. Liu, and J.-Z. Wang, *Effects of tau phosphorylation on proteasome activity*, FEBS Lett. **581**, 1521 (2007).
[86] S. J. Liu *et al.*, *Tau becomes a more favorable substrate for GSK-3 when it is prephosphorylated by PKA in rat brain*, J. Biol. Chem. **279**, 50078 (2004).
[87] H. Tak, M. M. Haque, M. J. Kim, J. H. Lee, J.-H. Baik, Y. Kim, D. J. Kim, R. Grailhe, and Y. K. Kim, *Bimolecular fluorescence complementation; lighting-up tau-tau interaction in living cells*, PloS one **8**, e81682 (2013).
[88] H.-H. Wang *et al.*, *Forskolin induces hyperphosphorylation of tau accompanied by cell cycle reactivation in primary hippocampal neurons*, Mol. Neurobiol. **55**, 696 (2018).
[89] Q. Tian *et al.*, *Biphasic effects of forskolin on tau phosphorylation and spatial memory in rats*, J. Alzheimer's Dis. **17**, 631 (2009).
[90] A. D. Poletayev, J. A. Dawson, M. S. Islam, and A. M. Lindenberg, *Defect-driven anomalous transport in fast-ion conducting solid electrolytes*, Nat. Mater. **21**, 1066 (2022).
[91] R. Zwanzig, *Memory effects in irreversible thermodynamics*, Phys. Rev. **124**, 983 (1961).
[92] H. Mori, *Transport, collective motion, and Brownian motion*, Prog. Theor. Phys. **33**, 423 (1965).

[93] H. Stehfest, *Algorithm 368: Numerical inversion of Laplace transforms [D5]*, Commun. ACM **13**, 47 (1970).
[94] S. Song, G.-S. Yang, S. J. Park, S. Hong, J.-H. Kim, and J. Sung, *Frequency spectrum of chemical fluctuation: A probe of reaction mechanism and dynamics*, PLoS Comput. Biol. **15**, e1007356 (2019).
[95] D. R. Cox, *Renewal theory* (Methuen, 1962).
[96] T. Miyaguchi, T. Akimoto, and E. Yamamoto, *Langevin equation with fluctuating diffusivity: A two-state model*, Phys. Rev. E **94**, 012109 (2016).
[97] J. Abate and P. P. Valkó, *Multi-precision Laplace transform inversion*, Int J Numer Methods Eng **60**, 979 (2004).
[98] J. Sung and R. J. Silbey, *Exact dynamics of a continuous time random walker in the presence of a boundary: Beyond the intuitive boundary condition approach*, Phys. Rev. Lett. **91**, 160601 (2003).

## Figures and Tables

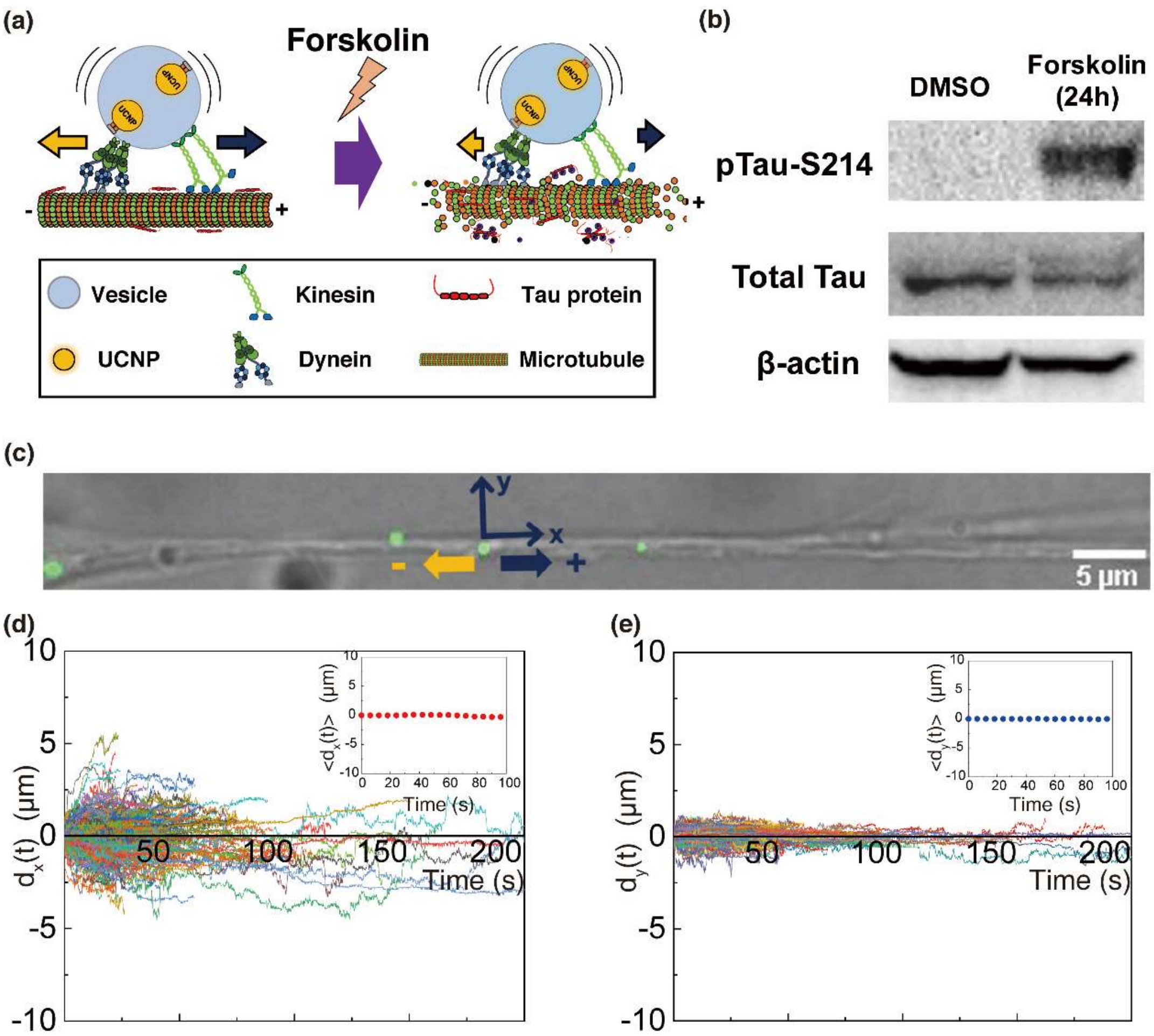

**FIG. 1. Trajectories of vesicle carried by motor protein multiplex on unstable microtubules in forskolin-treated cells.** (a) Schematic representation of vesicle-motor protein multiplex (VMPM) on the microtubule in normal cells (left) and in forskolin-treated cells (right). Under forskolin treatment, microtubules are destabilized by hyperphosphorylation of tau proteins and ensuing tau detachment from the microtubules. Both kinesins and dyneins simultaneously carry and transport a vesicle along the microtubules. Vesicle-encapsulated upconverting nanoparticles

(UCNP) are employed to track vesicle motion in our cell systems. (b) Western blotting image after forskolin treatment (20 μM for 24 hours) in SH-SY5Y cells. Tau-hyperphosphorylation at SER-214 site is induced selectively by forskolin treatment (see Fig. 4). (c) A snapshot of vesicle motion in live SH-SY5Y cells. The x-axis and y-axis directions designate the microtubule direction and microtubule-orthogonal direction, respectively. (d and e) Trajectories of VMPMs in the cells under hyperphosphorylation conditions. (insets) the mean displacement of VMPMs. In the hyperphosphorylated cells, trajectories of VMPMs have far different shapes from trajectories of VMPMs in normal cells (see Fig. 7). The average motion of VMPMs is not biased in any direction.

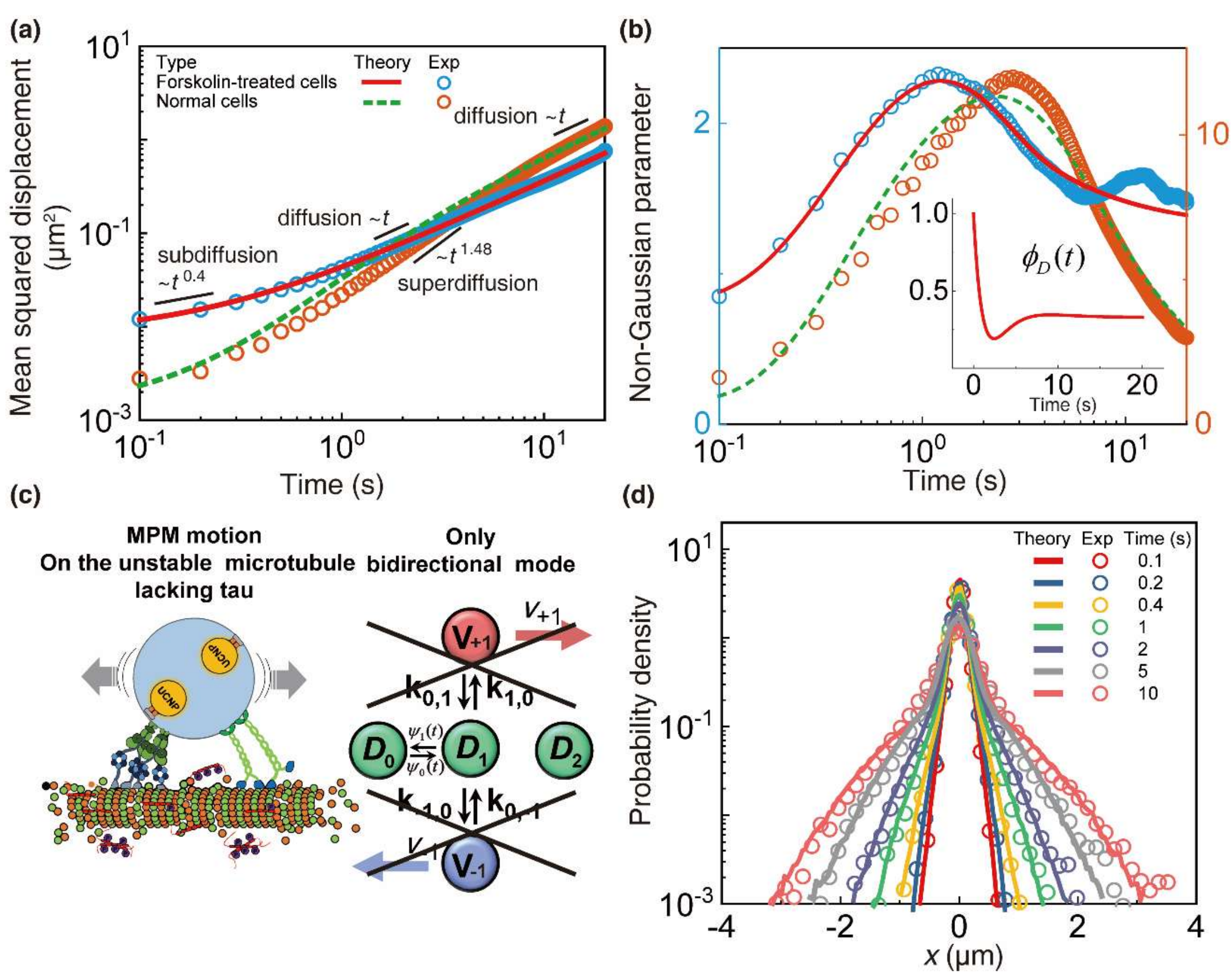

**FIG. 2. Mean square displacement, Non-Gaussian parameter, and Distribution of the vesicle motor protein multiplex's displacement along the microtubule.** (a and b) The mean square displacement (MSD) and the non-Gaussian parameter (NGP) of VMPM motion along the microtubule: (blue circles) experiment results of the forskolin-treated cells (red line) theoretical result of Eq. (1). (orange circles) experiment results of the normal cells (green dashed line) theoretical result of the multimode MPM model in Ref. [17]. In forskolin-treated cells, the MSD time-profile of VMPMs exhibits the direct transition from the initial sub-diffusion to ultimate diffusion without transient supper-diffusion that emerges in the MSD time-profile in normal cells; at short times, liberational motion of a vesicle bound to the MPM makes the dominant contribution to the MSD of the VMPM and at long times, diffusive motion of the MPM makes

the dominant contribution to the MSD, i.e., $\langle d_x^2\rangle \cong 2\langle D_\Gamma\rangle t$. (inset of B), The normalized time correlation function, $\phi_D(t)\left[\equiv\langle\delta D_\Gamma(t)\delta D_\Gamma(0)\rangle/\langle\delta D_\Gamma^2\rangle\right]$, of diffusion coefficient fluctuation of the MPM, extracted from the MSD and NGP data using Eq. (2) without assuming a particular model. Using the time-profile of $\phi_D(t)$, we construct the explicit model of the MPM motility fluctuation, shown in (c). In (b), the left-y axis and right y-axis represent the NGP value in the forskolin-treated cells and the NGP value in the normal cells, respectively. (c) Schematic representation of the MPM motion on an unstable microtubule lacking tau in the forskolin-treated cells. Unidirectional motion of the MPM is suppressed. MPMs only bidirectional random motion with a partially dynamic motility fluctuation. Our model of the MPM motion with motility fluctuation is schematically represented. $D_i$ designates the effective diffusion coefficient of the MPM in state $\Gamma_i$. A unified, quantitative explanation of our experimental results for the MSD and NGP time-profile is achieved when the waiting time distribution, $\psi_i(t)$ $i\in\{0,1\}$ of State $\Gamma_0$ or State $\Gamma_1$, is modeled by a gamma distribution (see Appendix B 7). (d) The VMPM displacement distribution in forskolin-treated cells: (circles) experiment results, (lines) theoretical predictions by our model optimized against the MSD and NGP data of VMPM motion in forskolin-treated cells.

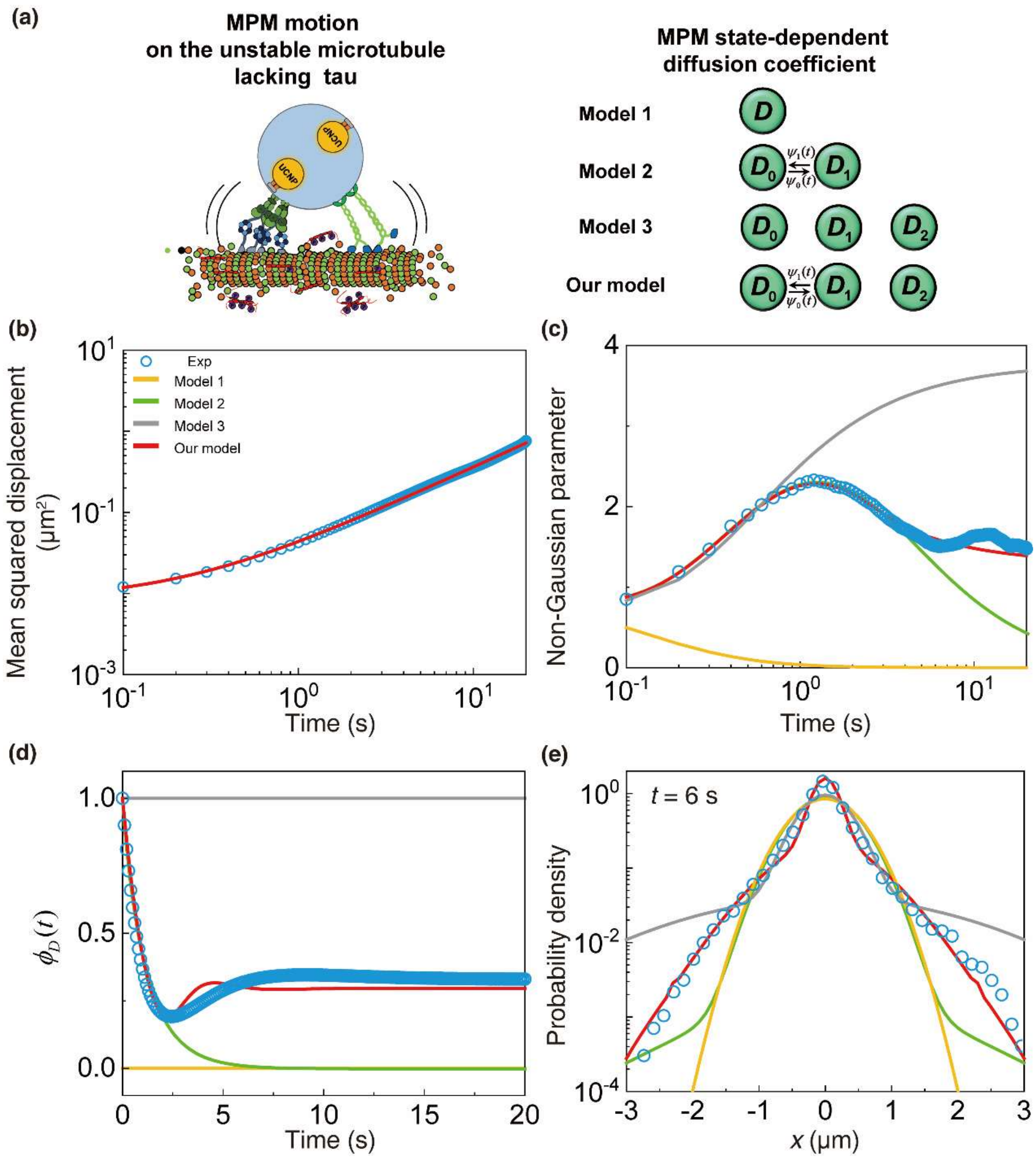

**FIG. 3. The MPM motility fluctuation dependent NGP time-profiles and the displace distribution of VMPMs.** (a) Various models of the MPM state dynamics. At the MPM state $\Gamma_i$, the MPM undergoes diffusive random motion with the MPM state-dependent diffusion coefficient, $D_i$ . $\psi_i(t)$ denotes the lifetime distribution of the MPM state $\Gamma_i$, modeled as a gamma distribution. See Table 2 for the parameter values used in the theoretical calculation for Model 1-3. For our model, the parameter values are listed in Table 1. (b and c) Time-profiles of the MSD and NGP of the VMPM. Theoretical results are calculated by Eq. (1) for all models. (d) Time correlation function of the diffusion coefficient fluctuation calculated for each model. For model 1, $\delta D = \phi_D(t) = 0$. For Model 2, the TCF is given by Eq. (4) with $p_2 = 0$. The values of other parameters are given in Table 2. For Model 3, $\phi_D(t) = 1$ because $\langle \delta D_{\mathbf{\Gamma}}(t) \delta D_{\mathbf{\Gamma}}(0) \rangle = \langle \delta D_{\mathbf{\Gamma}}^2 \rangle$ at all time *t*. The TCF of Model 3 is given by Eq. (4) with $G(\Gamma_i, t \,|\, \Gamma_j)$ replaced by $G(\Gamma_i, t \,|\, \Gamma_j) = \delta_{ij}$ with $\delta_{ij}$ denoting the Kronecker delta. The TCF of our model is given by Eq. (4). (e) The VMPM displacement distribution (VDD) at 6.0 sec. Theoretical results are calculated by numerical inversion of Eq. (B32)[97], where the expressions of $\hat{\hat{P}}(k,s)$ are given in Table 3 for Model 1, 2, and 3. For our model, $\hat{\hat{P}}(k,s)$ is given in Eq. (B16). For all models, $\tilde{f}(k)$ in Eq. (B32) is given by Eq. (B15). Model 1, 2, and 3 cannot explain our experimental results of the non-Gaussian parameter and the VDD regardless of the parameter values.

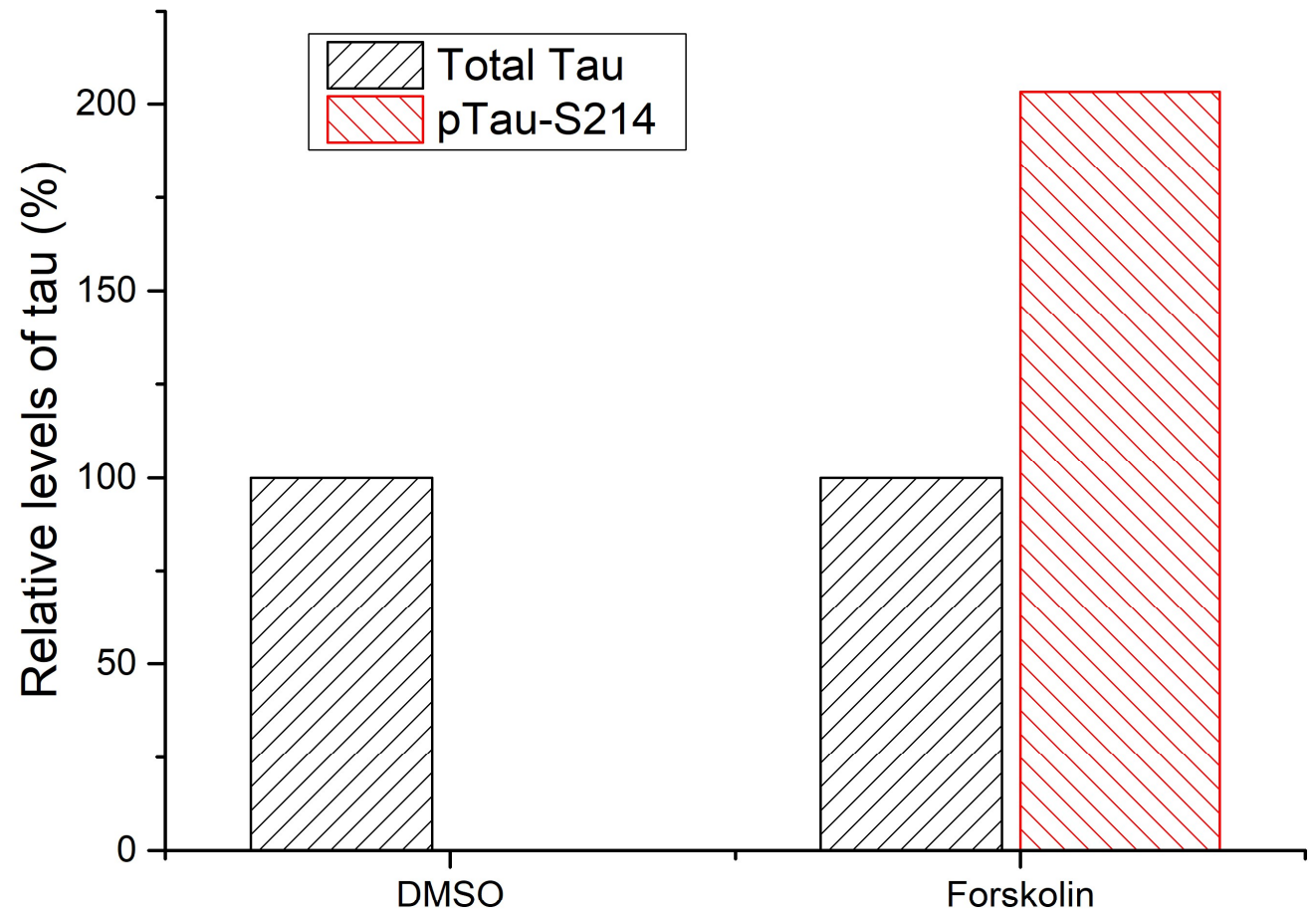

**FIG. 4**. **Relative levels of tau (%) in treated DMSO (total tau for normalization) and forskolin.** Cells were treated in 20 μM of forskolin for 24 hours. It is notable that nearly no signal from pTau-S214 solution with DMSO.

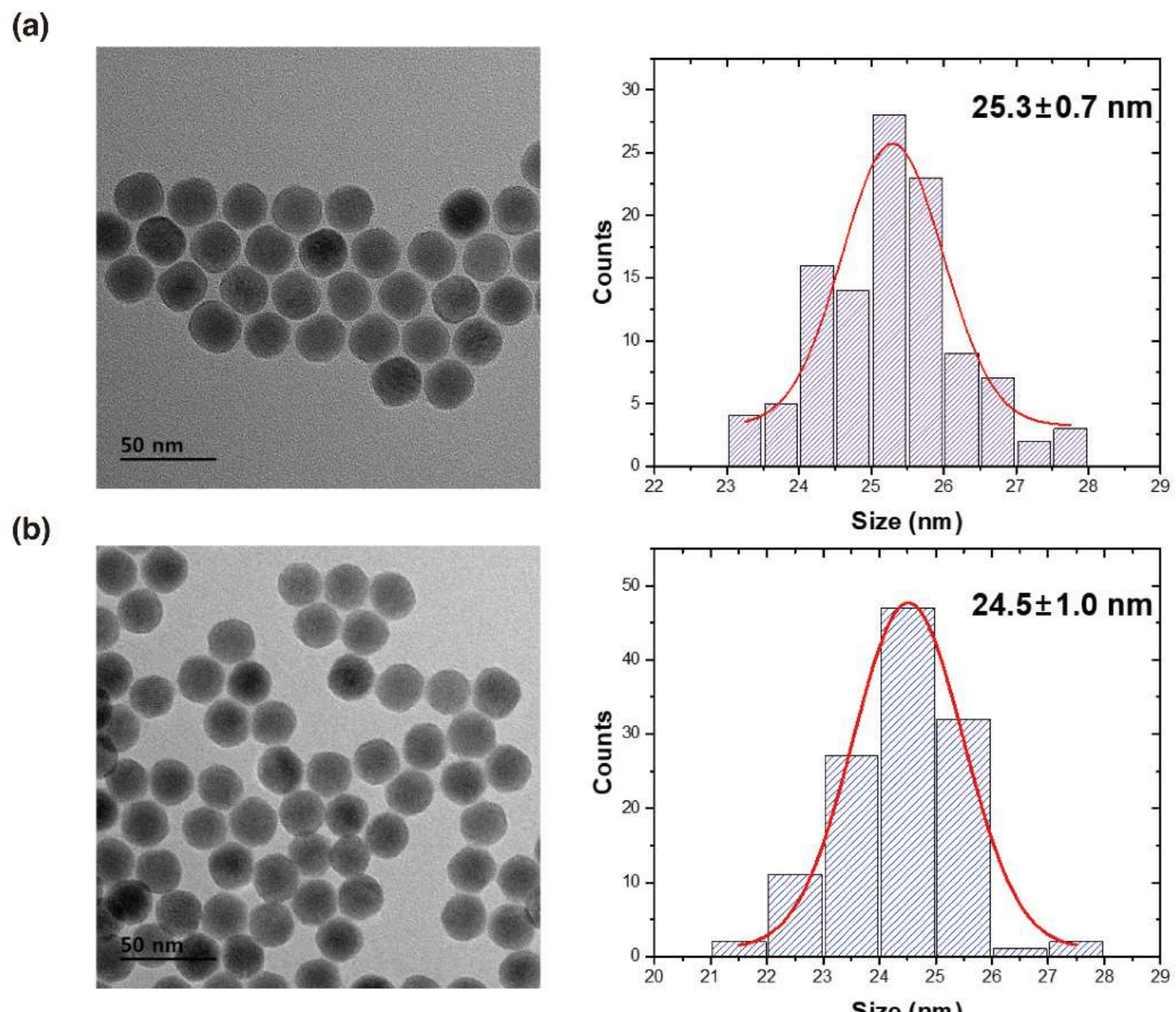

**FIG. 5. TEM image and size distribution of core/shell upconverting nanoparticles.** (a) TEM image and size distribution of core/shell $NaYF_4:Yb^{3+}$, $Er^{3+}$@$NaYF_4$ nanoparticles. The average size value of UCNPs which is calculated by Gaussian fitting is 25.3±0.7 nm. (b) TEM image and size distribution of core/shell $NaYF_4:Yb^{3+}$, $Er^{3+}$@$NaYF_4$@PAA nanoparticles. The average size value of UCNPs is 24.5±1.0 nm.

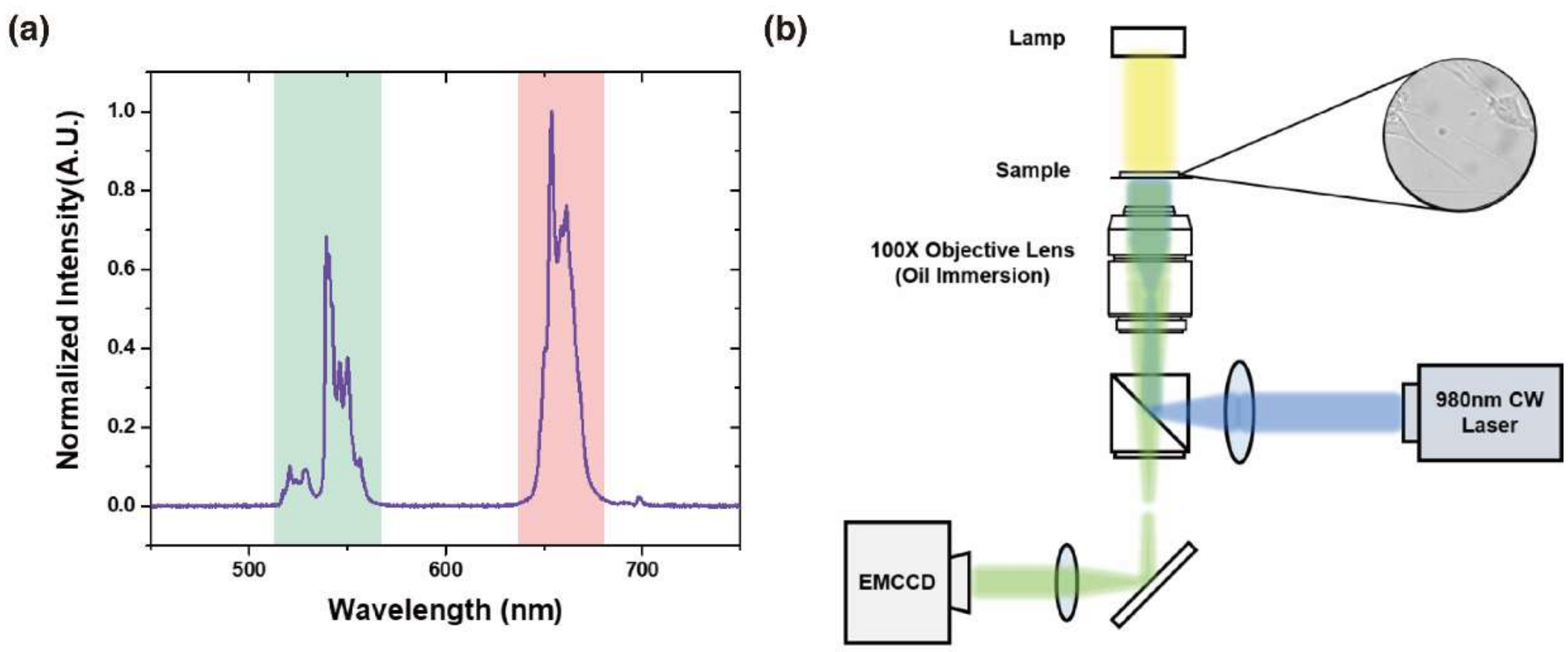

**FIG. 6.** (a) Emission spectrum of core/shell $NaYF_4$:$Yb^{3+}$, $Er^{3+}$@$NaYF_4$@PAA nanoparticles under 980-nm irradiation. Two main regions are observed in the emission spectrum (Green: 530 and 550 nm, Red: 650 nm). (b) Schematic diagram of the microscope setup for tracking vesicles containing upconverting nanoparticles (UCNPs) in the cell. Two light sources, lamp and 980nm CW laser, are used for cell imaging and excitation of UCNPs. Then, all images are detected by EMCCD.

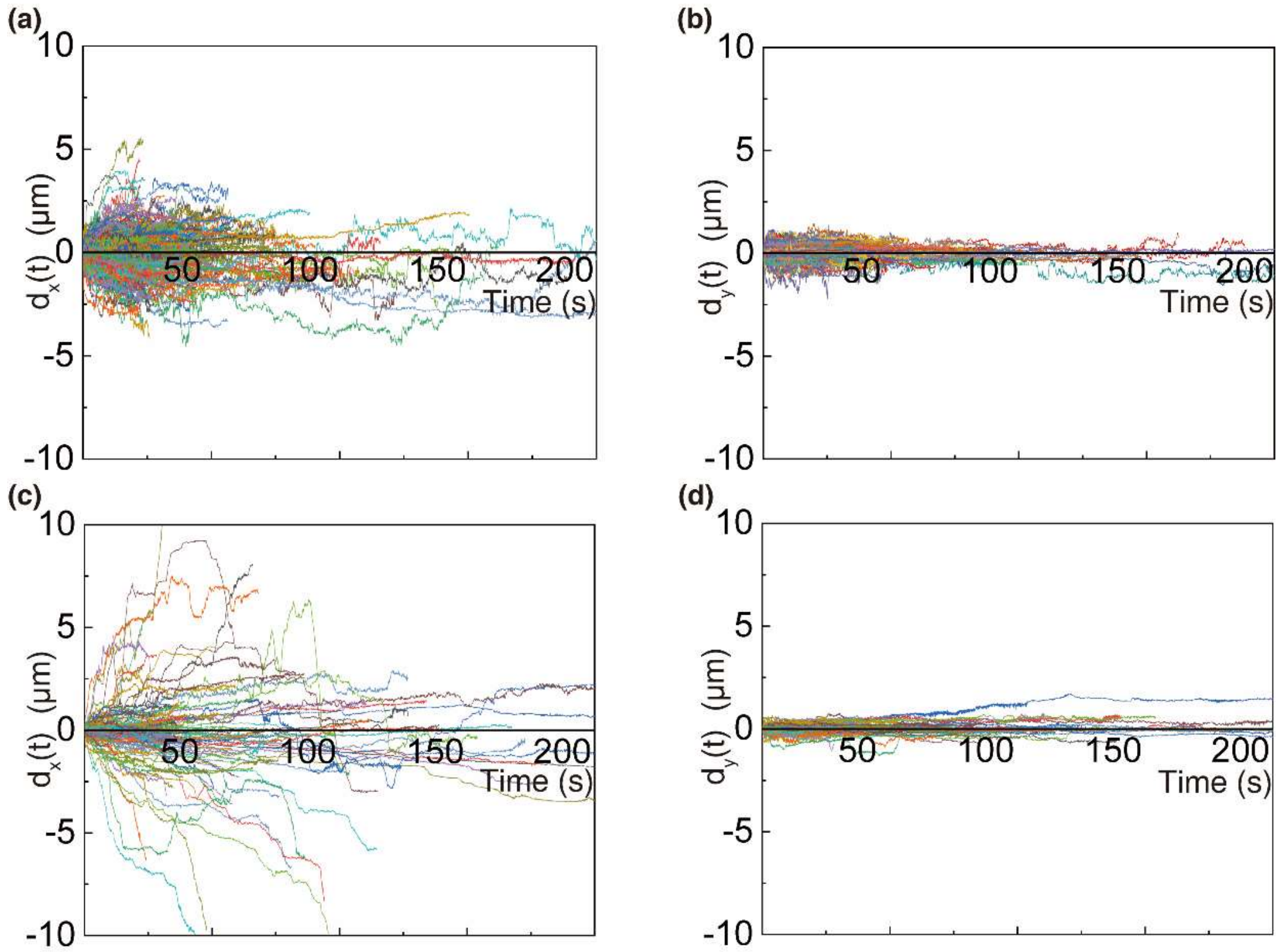

**FIG. 7. Trajectories of vesicle displacement.** (a and b) Trajectories of vesicle displacement in the cells under hyper-phosphorylation conditions. (c and d) Trajectories of vesicle displacement in normal cells. $d_{x(y)}$ in the y-axes represents the vesicle displacement in the microtubule (microtubule-orthogonal) direction. (c) and (b) are reproduced from Ref[17]

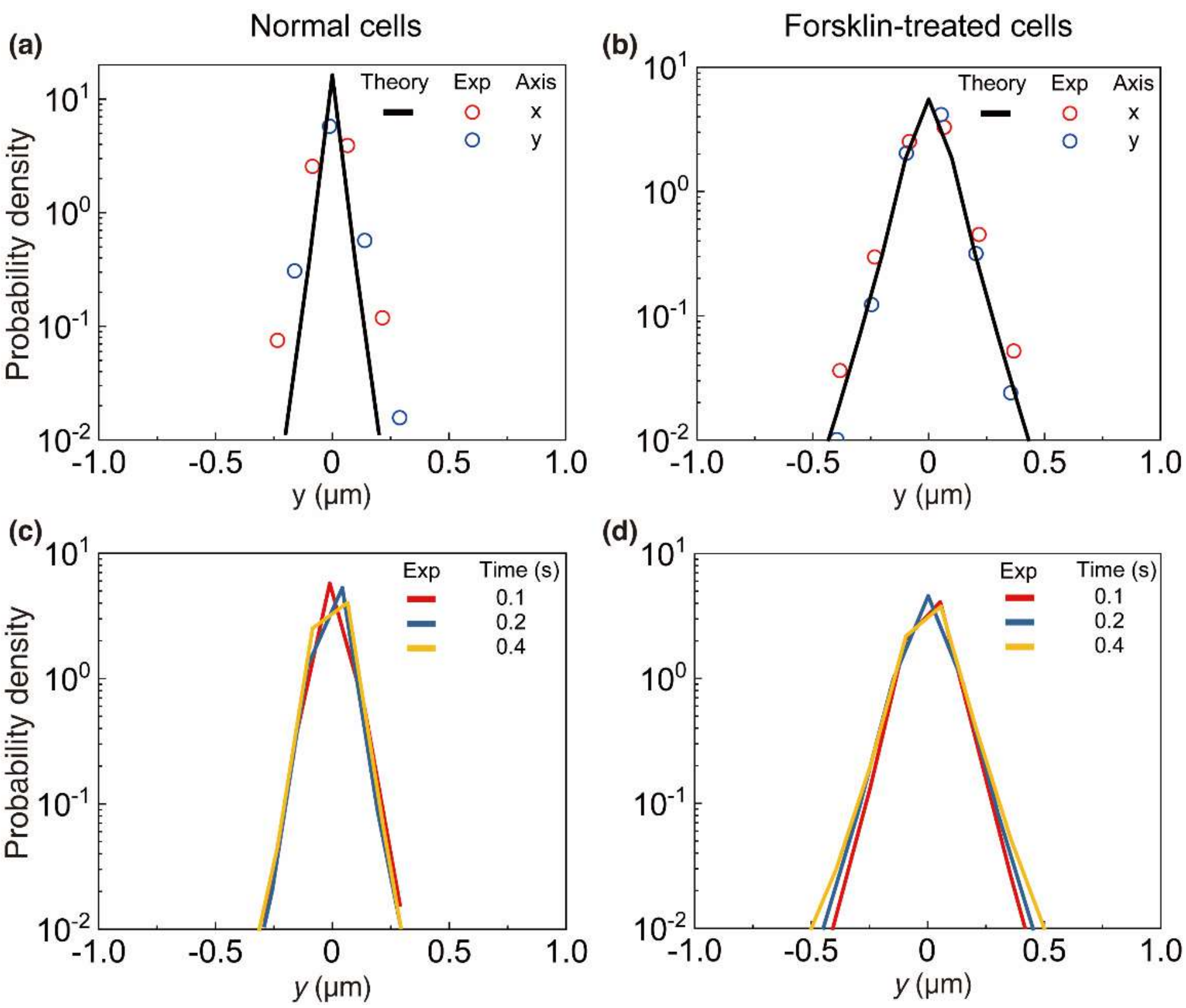

**FIG. 8. The vesicle displacement distribution at short times and in the microtubule-orthogonal direction in normal cells and forskolin-treated cells.** (a and b) Vesicle displacement distribution (VDD) at our experimental time resolution, 0.1 seconds (a) in normal cells (b) in forskolin-treated cells. (red circles) experiment results of the VDD in the microtubule direction, or the x-axis direction (blue circles) experiment results of the VDD in the microtubule-orthogonal direction, or the y-axis direction, (black line) theoretical result calculated by Eq. (B14) with $\kappa_0^{-1} = 1.99\times10^{-4}\,\mu\text{m}^2$ and $\sigma_q^2 = 5.30$ (normal cells)[17] and $\kappa_0^{-1} = 2.4\times10^{-3}\,\mu\text{m}^2$ and $\sigma_q^2 = 2.46$ (forskolin-treated cells). The liberational motion of the vesicle makes the dominant contribution to the VDD at short times where the MPM motion makes the negligible contribution. In both normal cells and forskolin-treated cells, the variance of the short-time VDD in the x-

direction is similar to the variance in the y-direction. The variance of the short-time VDD in forskolin-treated cells is greater than the variance in normal cells. (c and d) VDD in the microtubule-orthogonal direction at various times, (c) in normal cells and (d) in forskolin-treated cells. In both normal cells and forskolin-treated cells, the VDD in the microtubule-orthogonal direction hardly increases with time, in contrast with the VDD in the microtubule direction shown in Fig. 9

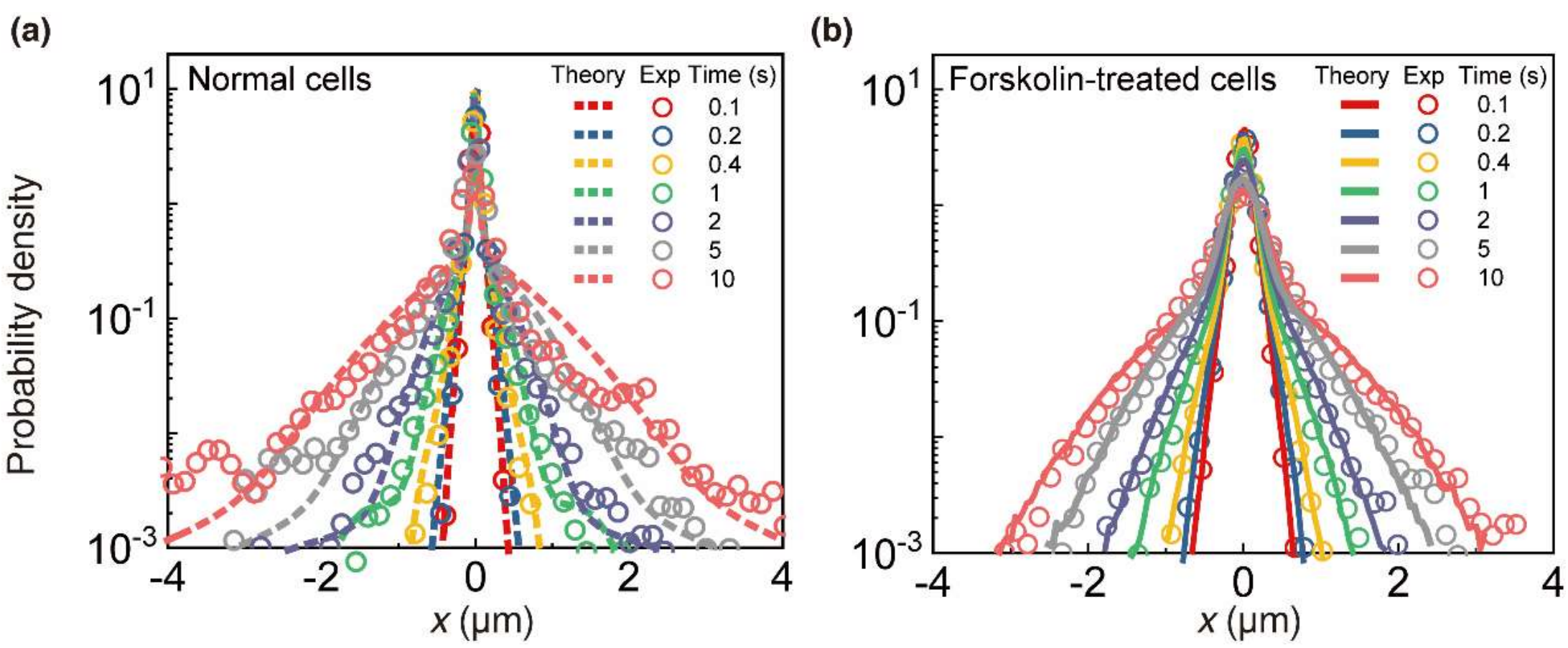

**FIG. 9. Comparison of vesicle displacement distribution between normal cells and forskolin-treated cells.** (circles) experimental results in normal cells and forskolin-treated cells (dashed lines) theoretical predictions by the multimode vesicle-MPM model in Ref[17] (lines) theoretical predictions by the vesicle-MPM model shown in Fig. 2(c), optimized against the MSD and NGP data.

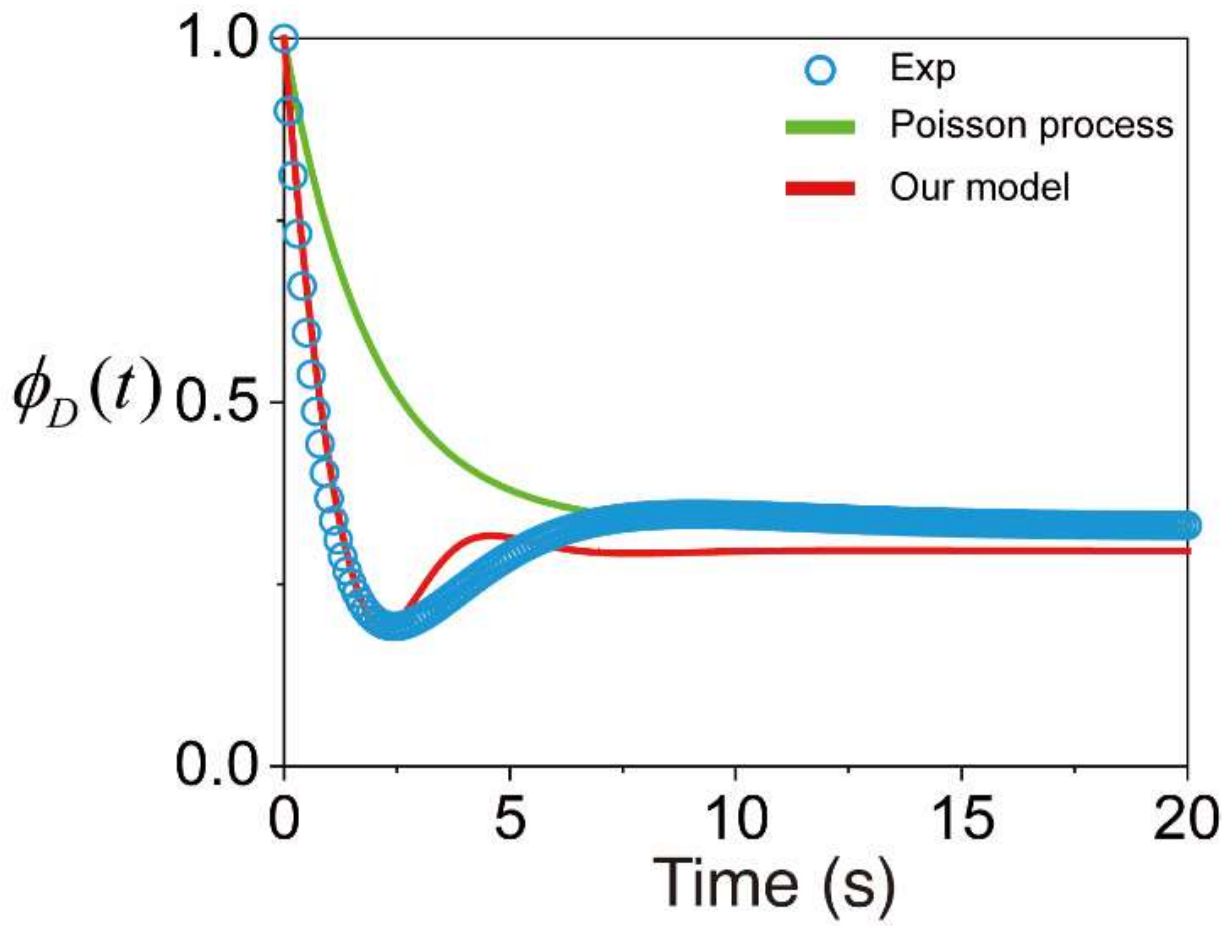

**FIG. 10. Time correlation function of the diffusion coefficient fluctuation $\phi_D(t)\left[=\langle\delta D_\Gamma(t)\delta D_\Gamma(0)\rangle\big/\langle\delta D_\Gamma^2\rangle\right]$.** (blue circles) TCF extracted from the experimental results of the MSD and NGP using Eq. (2) (red line) theoretical result of our MPM model, Eq. (4), with lifetime distributions $\psi_0(t)$ and $\psi_1(t)$ of MPM states, $\Gamma_0$ and $\Gamma_1$, being gamma distributions (see Fig. 11). The adjustable parameter values are given in Table 1. (green line) Theoretical result when the transitions between MPM states, $\Gamma_0$ and $\Gamma_1$, are the first-order kinetic processes. For this model, $\psi_0(t)$ and $\psi_1(t)$ of MPM states are given by simple exponential functions, i.e., $\psi_i(t)=k_i\exp(-k_i t)$, and $\phi_D(t)$ is given by $\phi_D(t)=\phi_D(\infty)+[1-\phi_D(\infty)]\exp[-t(k_1+k_0)]$. The best fitted parameter values are given by $k_1+k_2\cong 0.296$ and $\phi_D(\infty)\cong 0.526$.

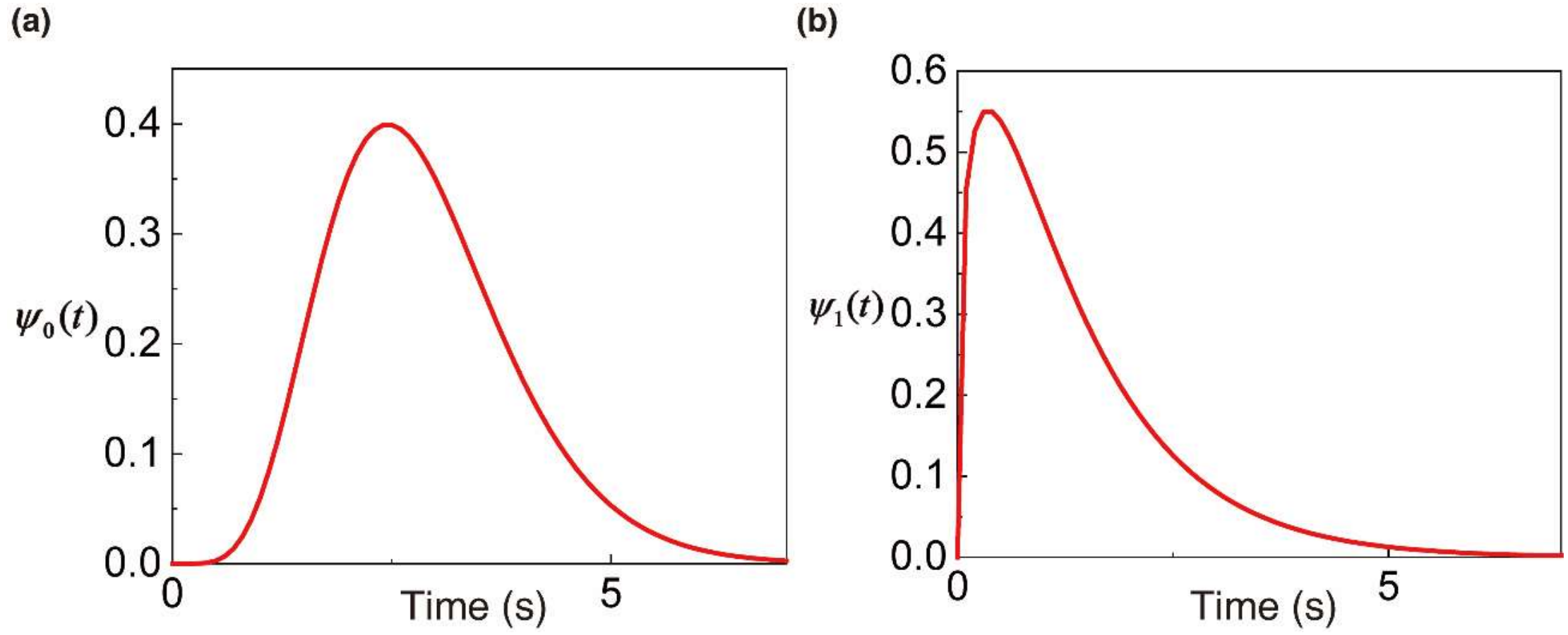

**FIG. 11. Lifetime distribution of the dynamic MPM states in our model, Eq. (3), extracted from our quantitative analysis of the MSD and NGP.** The lifetime distribution, $\psi_i(t)$, state $\Gamma_i$ is modeled as a gamma distribution, i.e., $\psi_i(t) = b_i^{-a_i} t^{a_i - 1} e^{-t/b_i} / \Gamma(a_i),\ i \in \{0,1\}$ with $\langle t_i \rangle = a_i b_i$ and $\langle \delta t_i^2 \rangle / \langle t_i \rangle^2 = a_i^{-1}$. The optimized parameter values are given in Table 1. These results are obtained by analyzing the experimental results for the MSD and NGP using Eq. (1).

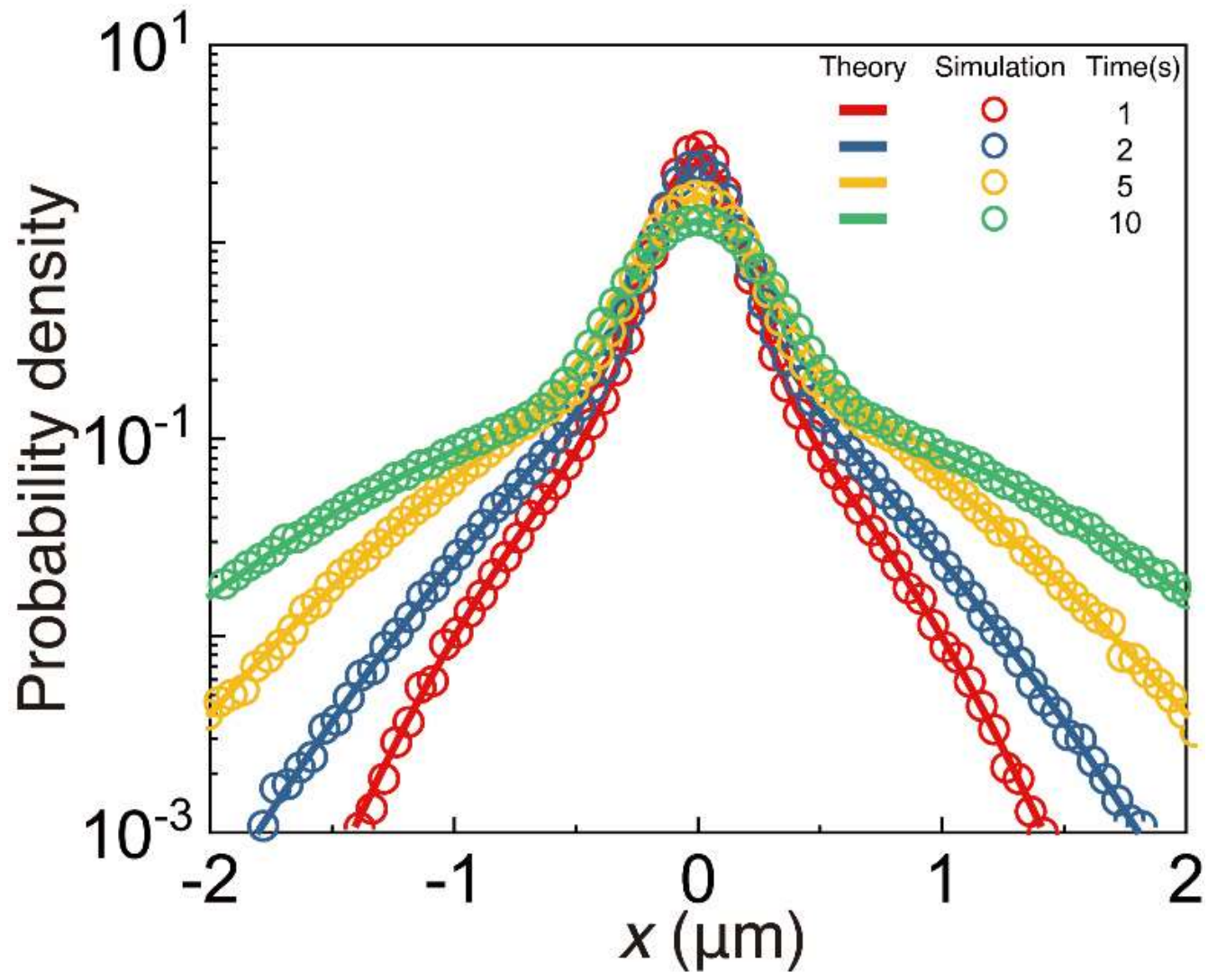

**FIG. 12. Comparison between theory and stochastic simulation for the vesicle displacement distribution. (**circles) simulation results for our optimized model; (lines) theoretical predictions by Eq. (B32) with the optimized parameter values in Table 1. The prediction of our model is in good agreement with experimental results for the vesicle displacement distribution as shown in Fig. 2(d) in the main text.

**Table 1. Optimized values of the adjustable parameters of the vesicle-motor protein multiplex model shown in Fig. 2(c).**

| Adjustable parameters | Values |
|---|---|
| $D_0$ | $8.95\times10^{-3}$ μm²/s |
| $D_1$ | $1.18\times10^{-1}$ μm²/s |
| $D_2$ | $2.12\times10^{-3}$ μm²/s |
| $p_2$ | 0.626 |
| $\langle t_0\rangle$ | 2.85 s |
| $\langle \delta t_0^2\rangle / \langle t_0\rangle^2$ | 0.14 |
| $\langle t_1\rangle$ | 1.34 s |
| $\langle \delta t_1^2\rangle / \langle t_1\rangle^2$ | 0.74 |

$D_0$, $D_1$, and $D_2$ designate the effective diffusion coefficient characterizing the bidirectional motion of the MPM at states, $\Gamma_0$, $\Gamma_1$, and $\Gamma_2$, respectively. $p_2$ denotes the probability that a vesicle is in State $\Gamma_2$. The probabilities, $p_0$ and $p_1$, of State $\Gamma_0$ and State $\Gamma_1$ are related to $p_2$ by $p_0 = (1-p_2)\langle t_0\rangle / (\langle t_0\rangle + \langle t_1\rangle) \cong 0.254$ and $p_1 = (1-p_2)\langle t_1\rangle / (\langle t_0\rangle + \langle t_1\rangle) \cong 0.120$, where $\langle t_i\rangle$ denotes the mean lifetime of State $\Gamma_i$. $\langle \delta t_i^2\rangle / \langle t_i\rangle^2$ denotes the relative variance of the lifetime distribution $\psi_i(t)$ of State $\Gamma_i$, which is modeled by a gamma distribution, i.e., $\psi_i(t) = b_i^{-a_i} t^{a_i-1} e^{-t/b_i} / \Gamma(a_i)$, $i \in \{0,1\}$ with $\langle t_i\rangle = a_i b_i$ and $\langle \delta t_i^2\rangle / \langle t_i\rangle^2 = a_i^{-1}$. The values of these parameters are extracted from the time-profiles of the mean square displacement and the non-Gaussian parameter of the vesicle-motor protein multiplex motion in forskolin-treated cells, shown in Figs. 2(a) and 2(b).

**Table 2. Parameter values used in the calculation for Model 1-3 shown in Fig. 3.**

| Adjustable parameters | Model 1 | Model 2 | Model 3 |
|---|---|---|---|
| $D$ | $0.0177\mu\mathrm{m}^2/\mathrm{s}$ | - | - |
| $D_0$ | - | $9.89\times10^{-1}\mu\mathrm{m}^2/\mathrm{s}$ | $7.98\times10^{-1}\mu\mathrm{m}^2/\mathrm{s}$ |
| $D_1$ | - | $1.63\times10^{-2}\mu\mathrm{m}^2/\mathrm{s}$ | $2.85\times10^{-1}\mu\mathrm{m}^2/\mathrm{s}$ |
| $D_2$ | - | - | $9.77\times10^{-3}\mu\mathrm{m}^2/\mathrm{s}$ |
| $p_0$ | - | $1.46\times10^{-3}$ | $8.99\times10^{-3}$ |
| $p_1$ | - | $9.98\times10^{-1}$ | $1.76\times10^{-1}$ |
| $p_2$ | - | - | $8.15\times10^{-1}$ |
| $\langle t_0\rangle$ | - | 1.23s | - |
| $\langle\delta t_0^2\rangle/\langle t_0\rangle^2$ | - | $8.91\times10^{-1}$ | - |
| $\langle t_1\rangle$ | - | $8.44\times10^{2}\mathrm{s}$ | - |
| $\langle\delta t_1^2\rangle/\langle t_1\rangle^2$ | - | $3.35\times10^{-2}$ | - |

In Model 1, the MPM has a single state. $D$ denotes the diffusion coefficient characterizing the bidirectional motion of the MPM at the state. In Model 2, the MPM has two different states, $\Gamma_0$ and $\Gamma_1$. $D_j$ denotes the diffusion coefficient of the MPM at state $\Gamma_j$. The lifetime distribution $\psi_j(t)$ of MPM state $\Gamma_j$ is modeled as a gamma distribution as in our model (see Table 1 caption). $\langle t_j\rangle$ and $\langle\delta t_j^2\rangle$ denote the mean and variance of the lifetime of the MPM state $\Gamma_j$. $p_j$ denotes the equilibrium probability of MPM state $\Gamma_j$. $p_0$ and $p_1$ are given by $p_0 = \langle t_0 \rangle / (\langle t_0 \rangle + \langle t_1 \rangle)$ and $p_1 = 1 - p_0$. In Model 3, the MPM has three different states, $\Gamma_0$, $\Gamma_1$, and $\Gamma_2$, and there is no transition among the three states. Model 1, 2, and 3 cannot explain our experimental results of the NGP time-profile and the VMPM displacement distribution regardless of parameter values.

**Table 3. Analytic expression of $\hat{\hat{P}}(k,s)$ for Model 1,2, and 3 shown in Fig. 3.**

| | |
|---|---|
| **Model 1** | $\dfrac{1}{s+Dk^2}$ |
| **Model 2** | $\dfrac{p_0}{s_0}+\dfrac{p_1}{s_1}-\dfrac{[1-\hat{\psi}_0(s_0)][1-\hat{\psi}_1(s_1)]}{(\langle t_0\rangle+\langle t_1\rangle)[1-\hat{\psi}_0(s_0)\hat{\psi}_1(s_1)]}\left(\dfrac{1}{s_0}-\dfrac{1}{s_1}\right)^2$ |
| **Model 3** | $\dfrac{p_0}{s_0}+\dfrac{p_1}{s_1}+\dfrac{p_2}{s_2}$ |

$\hat{\hat{P}}(k,s)$ designates the Fourier-Laplace transform of the probability distribution of the motor-protein-multiplex (MPM) displacement, i.e., $\hat{\hat{P}}(k,s)=\int_{-\infty}^{\infty} dR_x \exp(ikR_x)\int_0^{\infty} ds \exp(-st)P(R_x,t)$. $D$, $D_j$, $\langle t_0\rangle$ and $\langle t_1\rangle$ have the same meaning as in Table 2. $s_j$ denotes $s+D_jk^2$ with $s$ being the Laplace variable. $\hat{\psi}_j(s)$ denotes the Laplace transform of the lifetime distribution $\psi_j(t)$ of the MPM state $\Gamma_j$, i.e., $\hat{\psi}_j(s)=\int_0^{\infty} dte^{-st}\psi_j(t)$. $\psi_j(t)$ is modeled as a gamma distribution (see caption of Table 1).